\documentclass[a4paper,fleqn,usenatbib]{mnras}

\usepackage[T1]{fontenc}
\usepackage{ae,aecompl}

\usepackage{graphicx}	
\usepackage{amsmath}	
\usepackage{amssymb}	
\usepackage{subfigure}
\usepackage{multirow}
\usepackage{xcolor}
\hypersetup{draft}

\newcommand{\pcc}{\,{\rm cm}^{-3}}
\newcommand{\gcc}{\,{\rm g \, cm}^{-3}}
\newcommand{\pcs}{\,{\rm cm}^{-2}}

\newcommand{\kel}{\, {\rm K}}

\newcommand{\msun}{\, {\rm M}_\odot}
\newcommand{\nh}{n_{\rm H}}

\newcommand{\pc}{\, {\rm pc}}

\newcommand{\kpc}{\, {\rm kpc}}

\newcommand{\myr}{\, {\rm Myr}}
\newcommand{\kyr}{\, {\rm kyr}}
\newcommand{\ug}{\, {\rm \mu G}}

\newcommand{\kms}{\, {\rm km \, s^{-1}}}

\title[Line emission-SFR relationship]{NEATH V: the relationship between line emission from dense gas tracers and the star formation rate}

\author[Priestley et al.]{
  F. D. Priestley$^1$\thanks{Email: priestleyf@cardiff.ac.uk}, P. C. Clark$^1$, S. C. O. Glover$^2$, S. E. Ragan$^1$, S. K. Stuber$^3$, M. T. Cusack$^1$, \newauthor O. Feh\'{e}r$^1$, R. S. Klessen$^{2,4}$
\\
$^{1}$School of Physics and Astronomy, Cardiff University, Queen's Buildings, The Parade, Cardiff CF24 3AA, UK \\
$^{2}$Universit\"{a}t Heidelberg, Zentrum f\"{u}r Astronomie, Institut f\"{u}r Theoretische Astrophysik, Albert-Ueberle-Stra{\ss}e 2, D-69120 Heidelberg, Germany \\
$^{3}$National Astronomical Observatory of Japan, National Institute of Natural Sciences, 2-21-1 Osawa, Mitaka, Tokyo 181-8588, Japan \\
$^{4}$Universit\"{a}t Heidelberg, Interdisziplin\"{a}res Zentrum f\"{u}r Wissenschaftliches Rechnen, Im Neuenheimer Feld 205, D-69120 Heidelberg, Germany \\
}

\date{Accepted XXX. Received YYY; in original form ZZZ}

\pubyear{2025}

\begin{document}
\label{firstpage}
\pagerange{\pageref{firstpage}--\pageref{lastpage}}
\maketitle

\begin{abstract}

The Gao-Solomon relationship between the luminosity of the HCN $J=1-0$ line and the star formation rate (SFR) is observed to remain close to linear over scales ranging from individual star-forming clumps to entire galaxies. This is widely interpreted as the HCN line tracing the reservoir of dense gas directly associated with star formation. However, resolved observations of nearby molecular clouds have demonstrated that the threshold density above which star formation occurs is significantly higher than that of the gas traced by HCN emission. We perform radiative transfer modelling of molecular line emission from simulated clouds, based on magnetohydrodynamic simulations with realistic gas and dust thermodynamics and a time-dependent treatment of the molecular abundances. We find no correlation between HCN emission and the SFR in the simulations: the HCN line remains almost constant in brightness over several orders of magnitude in SFR. The N$_2$H$^+$ $J=1-0$ line correlates positively with the SFR, but weakly, and with a substantial dependence on environmental conditions. The strongest correlation between line emission and physical cloud properties is between the N$_2$H$^+$/HCN ratio and the dense gas fraction, which is close to linear. We argue that the observed HCN-SFR correlation on extragalactic scales is a result of each measurement integrating over many individual molecular clouds, which, on average, possess the same mass fraction of dense, star-forming gas. The HCN line does not directly trace this reservoir for star formation.

\end{abstract}

\section{Introduction}

On extragalactic scales, the measured star formation rate (SFR) is closely related to the surface density of the interstellar medium (ISM), typically expressed as a power-law with a slope $\sim 1.4$ \citep{kennicutt1998}. When only considering the molecular ISM, as determined from CO line observations, the slope of the relationship may be close to unity (\citealt{bigiel2008,leroy2008}; although see \citealt{shetty2014}). This linear relationship between molecular gas mass and the SFR has led to the widespread assumption that molecular gas acts as the reservoir for star formation, motivating much work on determining the molecular fraction of the ISM \citep[e.g.][]{krumholz2011}, and various subgrid models of star formation in galaxy-scale simulations \citep{naab2017}.

In the Milky Way, most of the molecular gas appears to be forming stars either inefficiently, or not at all \citep{zuckerman1974}. On resolved molecular cloud scales, the SFR is instead correlated with the mass of high column density gas, with a star formation `threshold' of around $10^{22} \pcs$, while the total mass of molecular gas has almost no bearing on the cloud's SFR \citep{lada2010,heiderman2010}. On extragalactic scales, this is assumed to be the basis of the linear Gao-Solomon relationship between the SFR and the luminosity of the HCN $J = 1-0$ line \citep{gao2004}: the HCN line traces the mass of the dense gas directly associated with star formation. The linear relationship has been proposed to extend down to the scales of individual star-forming clumps in the Milky Way \citep{wu2005,wu2010}, making it a universal law of star formation, although with substantial scatter \citep{neumann2025} and potential systematic offsets between clumps and galaxies \citep{stephens2016}.

Although widely accepted, the proposed direct connection between HCN emission and star-forming gas is demonstrably incorrect on resolved cloud scales. Local molecular clouds show widespread HCN emission down to columns well below the star formation threshold \citep{pety2017,kauffmann2017,evans2020,patra2022}, and its integrated intensity is linearly correlated with the total column \citep{tafalla2021,tafalla2023}, not specifically that of high-density material. Radiative trapping effects make HCN and other supposed dense gas tracers emit well below their critical densities \citep{shirley2015}, and its high abundance in moderate-density gas \citep{priestley2023b} means that its line emission is dominated by this material \citep{jones2023,priestley2024}, rather than the less-prevalent high-density cloud component. On local cloud scales, the only commonly-observed molecule found to specifically trace the highest-column material is N$_2$H$^+$ \citep{pety2017,kauffmann2017,tafalla2021,yun2021}. This has led to significant interest in its potential as a superior dense gas tracer to HCN on extragalactic scales \citep{jimenez2023,stuber2023,feher2024}, albeit a challenging one to detect due to its intrinsic faintness.

In \cite{priestley2023c}, we used simulations of the chemical evolution of molecular clouds to show that N$_2$H$^+$ is an excellent tracer not just of dense gas, but specifically the cloud material linked to ongoing star formation. This is due to its rapid destruction by {both free electrons and gas-phase CO, only existing in detectable quantities in cold, dense, well-shielded regions where the ionisation fraction is low and CO is strongly depleted onto grain surfaces} \citep{caselli2002}. The freeze-out timescale at a density of $10^4 \pcc$ (where CO depletion becomes significant; \citealt{tafalla2002,priestley2023b}) is $\sim 0.4 \myr$, coincidentally almost identical to the free-fall timescale at this density: the conditions necessary for detectable levels of N$_2$H$^+$ emission are identical to those describing gravitationally-collapsing high-density gas. No other commonly-observed dense gas tracer shares this behaviour.

Previous efforts to understand the relationship between molecular line emission and the SFR have typically neglected at least one important aspect of the problem, such as by imposing the molecular abundances \citep{jones2023}, the density structure \citep{bemis2024}, or the column-intensity relationship \citep{zakardjian2025} on the model. In this paper, we use radiative transfer modelling to produce synthetic molecular line observations of simulated star-forming clouds, including a full treatment of ISM thermodynamics and the time-dependent chemical evolution via the NEATH (Non-Equilibrium Abundances Treated Holistically; \citealt{priestley2023b}) framework. With the SFR of the simulated clouds known, we explore how observed extragalactic correlations between star formation and line emission emerge from the underlying cloud-scale properties, finding that they are likely a result of averaging over large ensembles of fundamentally-similar molecular clouds.

\section{Method}

\begin{table}
  \centering
  \caption{Simulation parameters, duration, and the number and total mass of sink particles formed at the simulation endpoint.}
  \begin{tabular}{cccccc}
    \hline
    $\gamma$ & $G / {\rm G_0}$ & $\zeta / {\rm s^{-1}}$ & $t_{\rm end} / \myr$ & $N_{\rm sink}$ & $M_{\rm sink} / \msun$ \\
    \hline
    $1$ & $1.7$ & $10^{-16}$ & $5.53$ & $105$ & $102$ \\
    $30$ & $51$ & $3 \times 10^{-15}$ & $7.31$ & $76$ & $158$ \\
    \hline
  \end{tabular}
  \label{tab:arepo}
\end{table}

\begin{table}
  \centering
  \caption{Elemental abundances used in the chemical modelling.}
  \begin{tabular}{ccccc}
    \hline
    Element & Abundance & & Element & Abundance \\
    \hline
    C & $1.4 \times 10^{-4}$ & & S & $1.2 \times 10^{-7}$ \\
    N & $7.6 \times 10^{-5}$ & & Si & $1.5 \times 10^{-7}$ \\
    O & $3.2 \times 10^{-4}$ & & Mg & $1.4 \times 10^{-7}$ \\
    \hline
  \end{tabular}
  \label{tab:abun}
\end{table}

\begin{table}
  \centering
  \caption{Collisional partners for the line radiative transfer and sources of collisional rate data.}
  \begin{tabular}{ccc}
    \hline
    Molecule & Partners & Reference \\
    \hline
    $^{12}$CO & p-H$_2$, o-H$_2$ & \citet{yang2010} \\
    \quad \\
    \multirow{2}{*}{HCN} & H$_2$ & \citet{dumouchel2010} \\
    & e$^-$ & \citet{faure2007} \\
    \quad \\
    N$_2$H$^+$ & H$_2$ & \citet{flower1999} \\
    HCO$^+$ & p-H$_2$, o-H$_2$ & \citet{denis2020} \\
    \hline
  \end{tabular}
  \label{tab:lamda}
\end{table}

\begin{figure*}
  \centering
  \includegraphics[width=\textwidth]{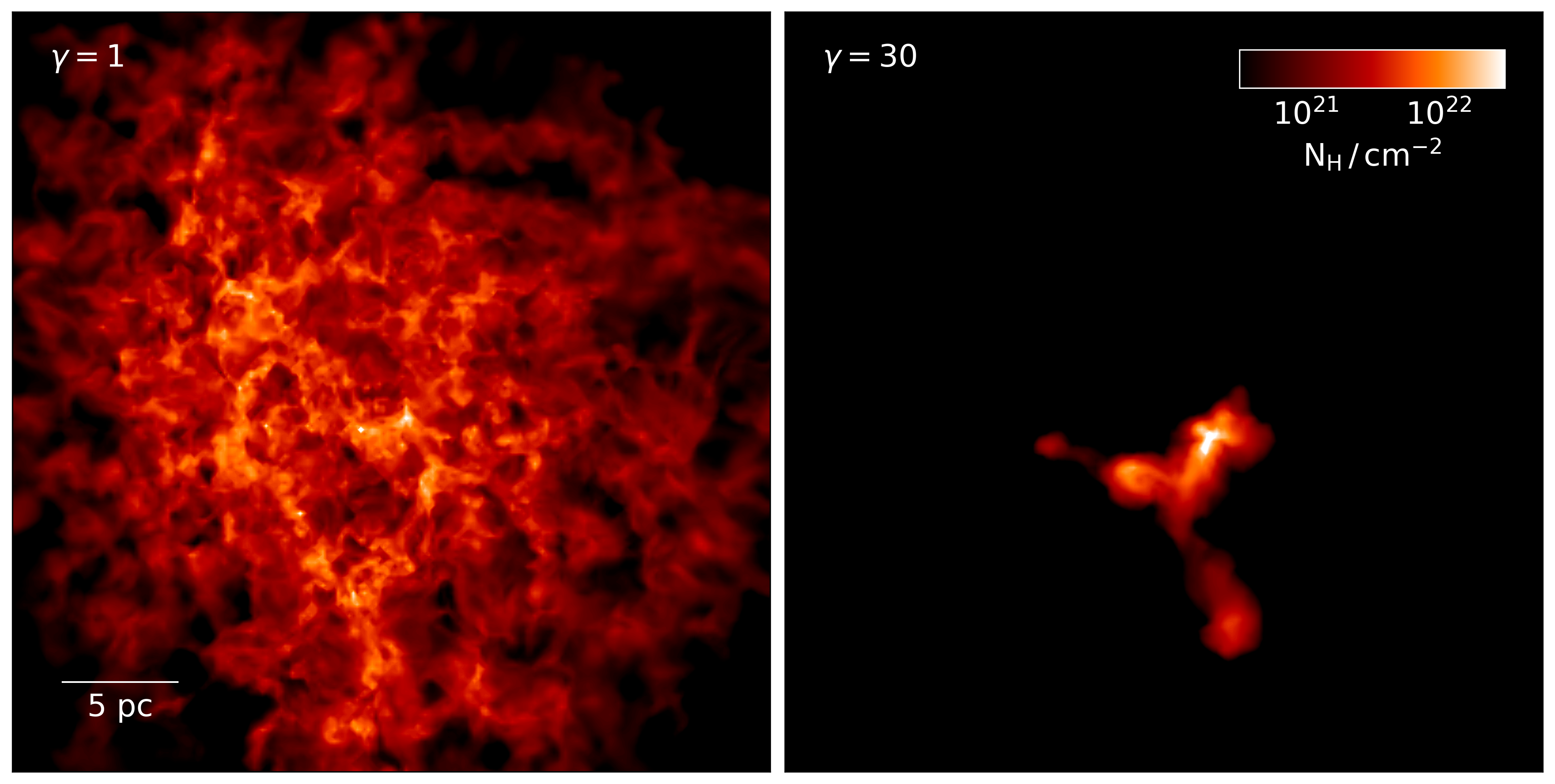}
  \caption{Total column density maps seen face-on to the collision at the endpoint of the $\gamma = 1$ (left) and $30$ (right) simulations.}
  \label{fig:facecol}
\end{figure*}

\begin{figure}
  \centering
  \includegraphics[width=\columnwidth]{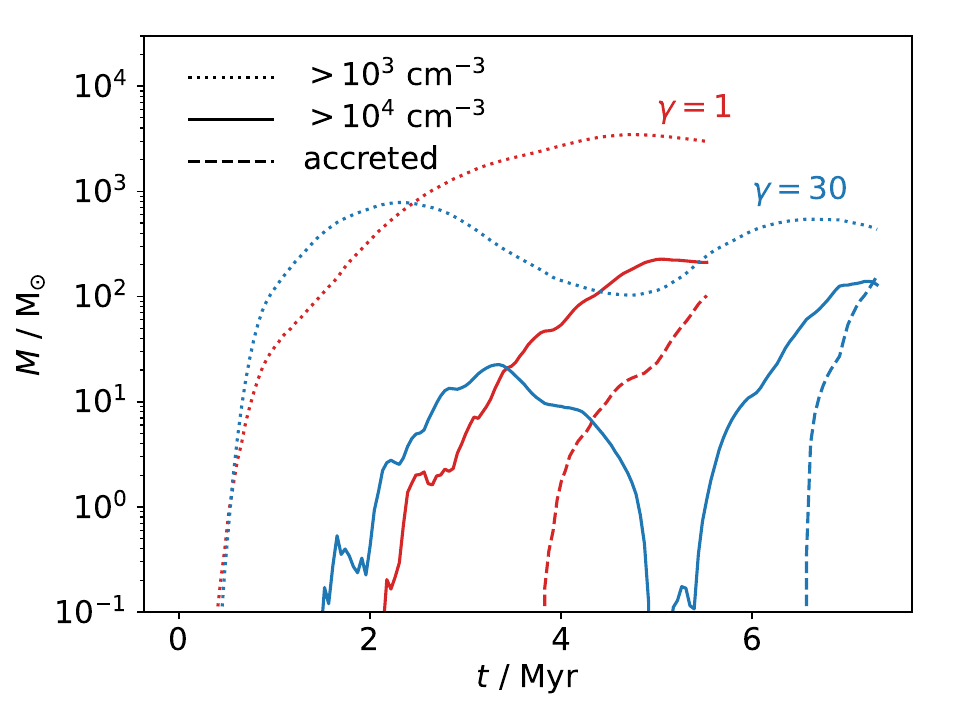}
  \caption{Evolution of the mass of sink particles (dashed lines) and the gas mass above a density of $10^3 \pcc$ (dotted lines) and $10^4 \pcc$ (solid lines) for the $\gamma = 1$ (red) and $30$ (blue) simulations.}
  \label{fig:mass}
\end{figure}

\begin{figure*}
  \centering
  \includegraphics[width=\columnwidth]{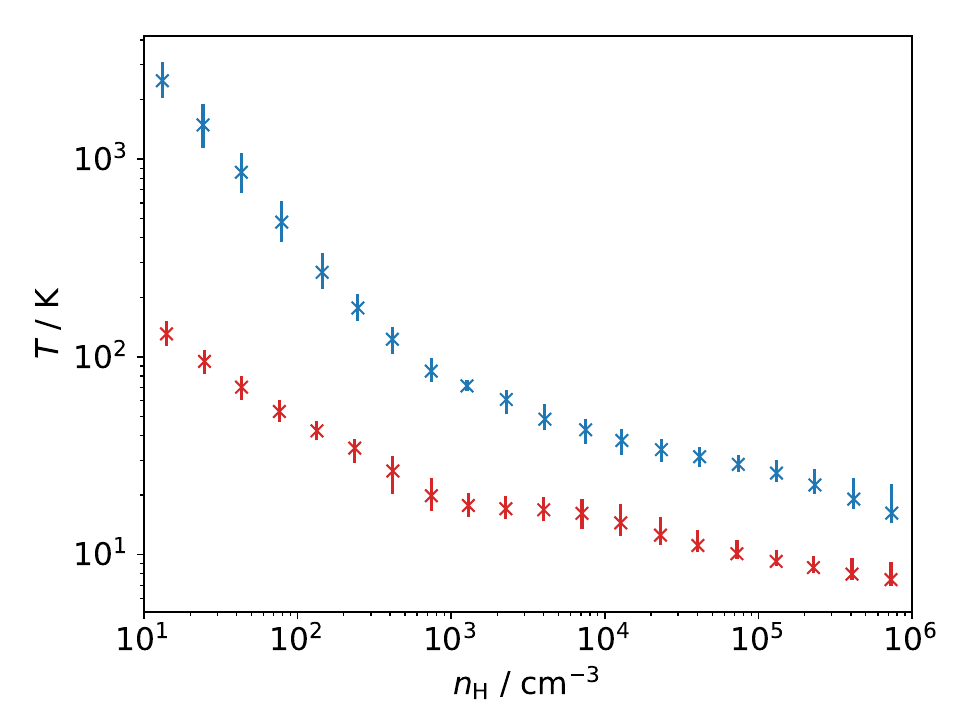}
  \includegraphics[width=\columnwidth]{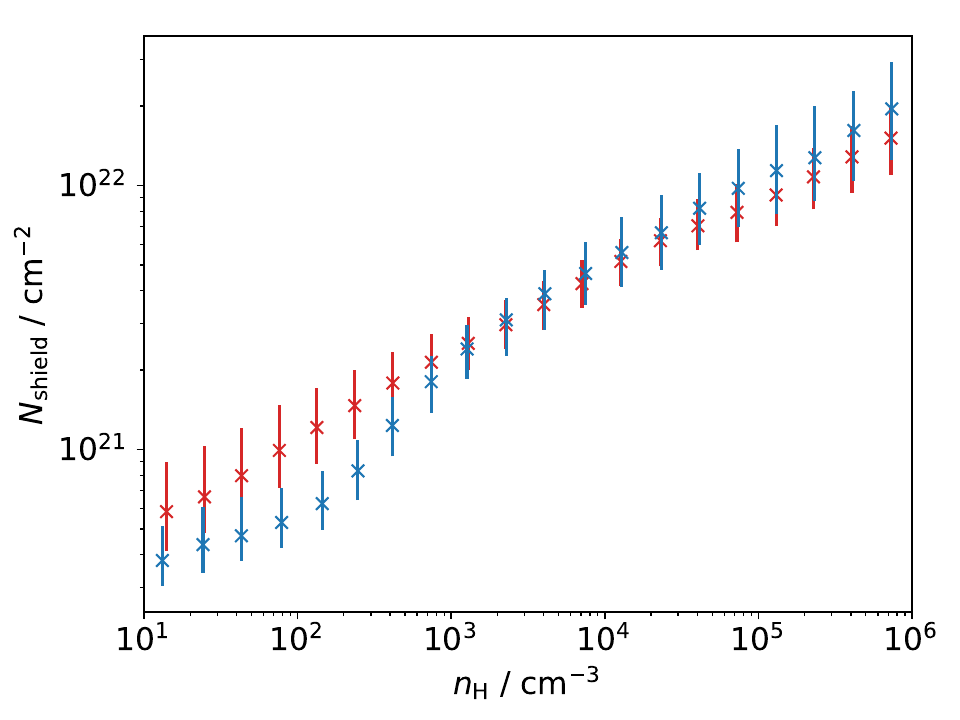}
  \caption{Average gas temperatures (left) and effective shielding column densities (right) as a function of volume density for the $\gamma = 1$ (red) and $30$ (blue) simulations. The median values are shown as crosses, with the bars indicating the 16th/84th percentiles.}
  \label{fig:physprop}
\end{figure*}

\begin{figure*}
  \centering
  \includegraphics[width=\columnwidth]{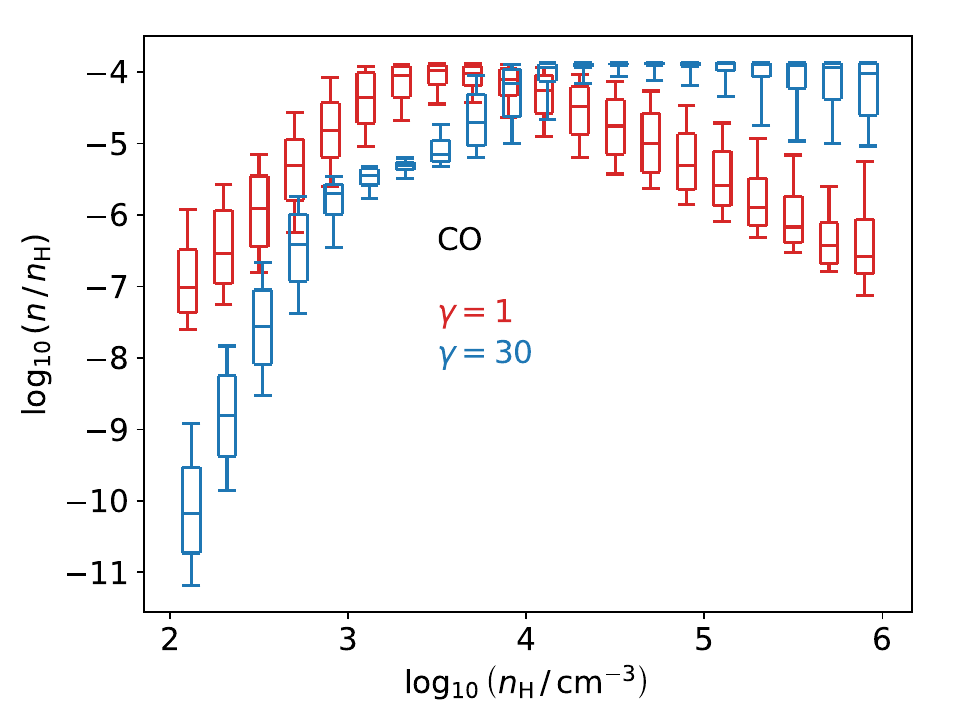}
  \includegraphics[width=\columnwidth]{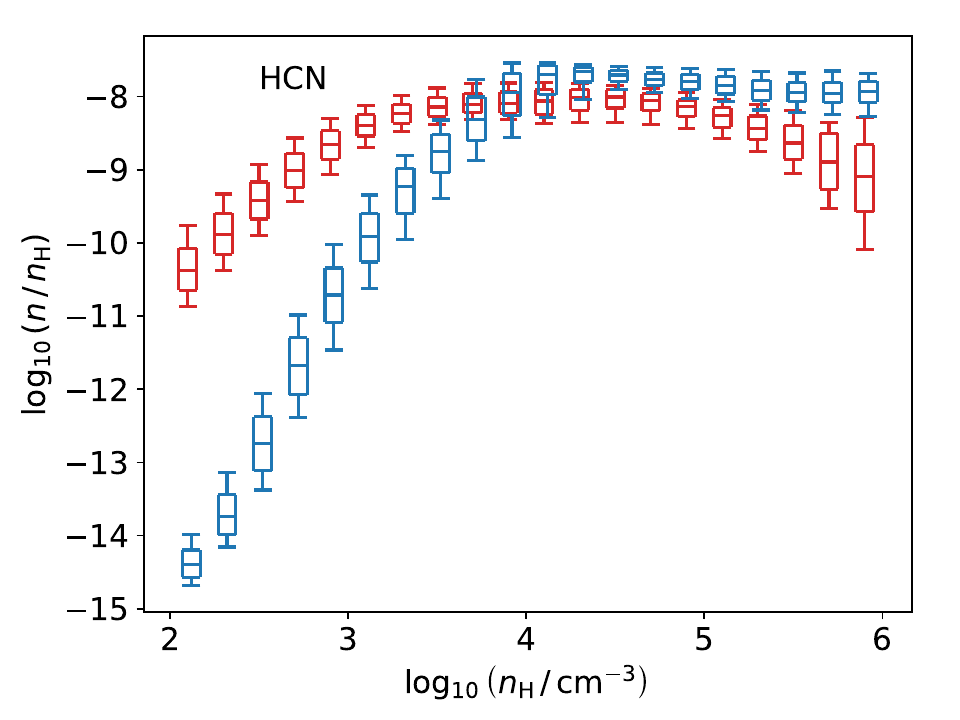}\\
  \includegraphics[width=\columnwidth]{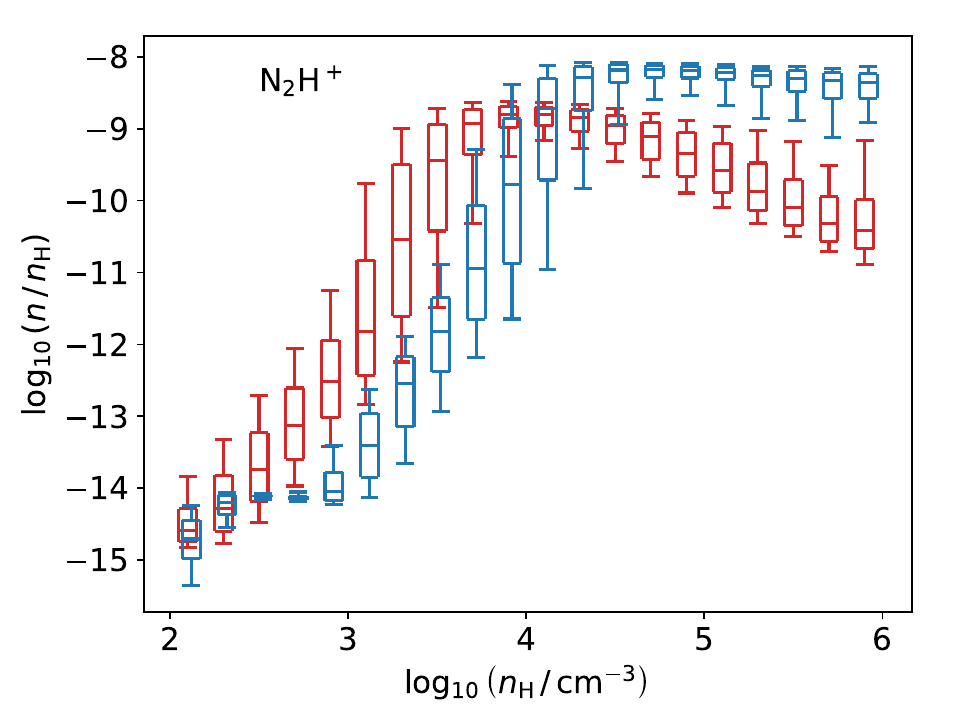}
  \includegraphics[width=\columnwidth]{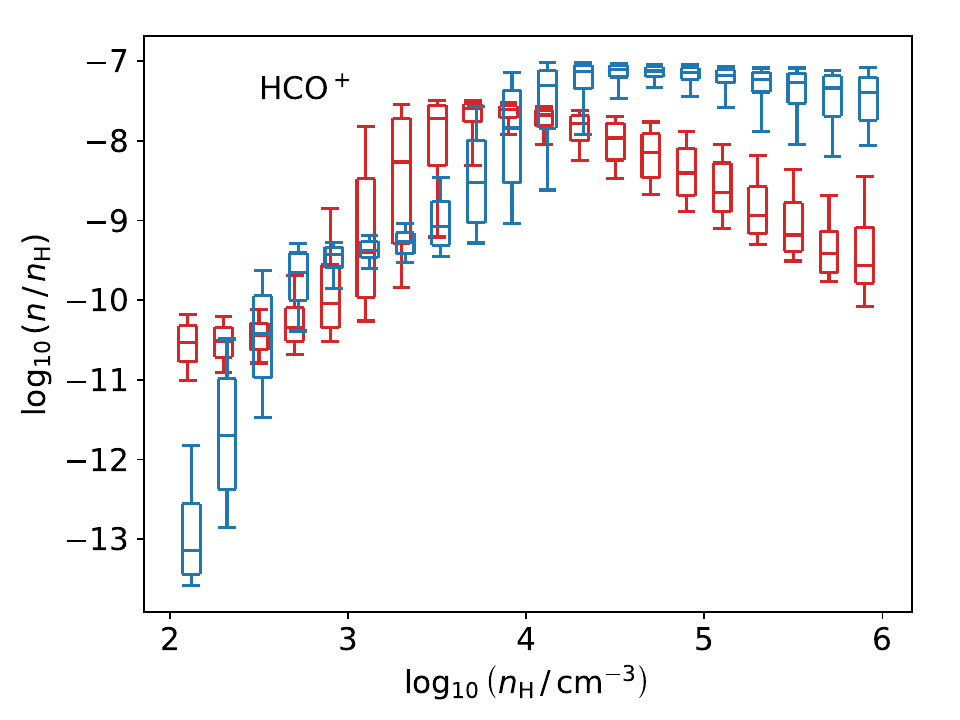}
  \caption{Molecular abundances with respect to hydrogen nuclei versus volume density for the $\gamma = 1$ (red) and $30$ (blue) simulations. Boxes show the median abundance and the 25th/75th percentiles, whiskers the 10th/90th.}
  \label{fig:chem}
\end{figure*}

\subsection{Cloud simulations}

We perform magnetohydrodynamic (MHD) simulations of star-forming clouds using the {\sc arepo} moving-mesh code \citep{springel2010,pakmor2011}, with the inclusion of the ISM thermochemical model described in \citet{glover2007} and \citet{glover2012b}. The H$_2$ and CO chemistry is followed on-the-fly using the \citet{gong2017} reaction network, with some alterations as described in \citet{hunter2023}. Shielding column densities are calculated self-consistently using the TreeCol algorithm \citep{clark2012a}. Sink particles are introduced according to the criteria in \citet{tress2020}, with a threshold density of $2 \times 10^{-16} \gcc$ and a formation radius of $9 \times 10^{-4} \pc$. We track the evolution of individual parcels of gas using Monte Carlo tracer particles \citep{genel2013}.

Our simulation setup is a head-on collision between two spherical gas clouds, with the magnetic field oriented along the collision axis. The initial cloud masses and radii are $10^4 \msun$ and $19 \pc$ respectively, giving an initial number density of hydrogen nuclei $\nh = 10 \pcc$. We assume the clouds are initially atomic, with gas and dust temperatures of $300 \kel$ and $15 \kel$ respectively. Each cloud is given a bulk velocity of $7 \kms$ toward the other, for a net collision velocity of $14 \kms$, and a virialised turbulent velocity field with three-dimensional dispersion $0.95 \kms$. The magnetic field strength is $3 \ug$, oriented along the collision axis, giving a mass-to-flux ratio of $2.4$ times the critical value for collapse \citep{mouschovias1976}.

We assume carbon and oxygen abundances of $1.4 \times 10^{-4}$ and $3.2 \times 10^{-4}$ from \citet{sembach2000}, and a `metal' abundance of $10^{-7}$, taken to be sillicon in the \citet{gong2017} network. We take standard Solar neighbourhood environmental conditions to be characterised by an ultraviolet (UV) radiation field of $G = 1.7 \, {\rm G_0}$ \citep{draine1978}, where ${\rm G_0}$ is the \citet{habing1968} field, and a cosmic ray ionisation rate (CRIR) for atomic hydrogen of $\zeta = 10^{-16} \, {\rm s^{-1}}$ \citep{indriolo2012}. We also investigate a more extreme environment, with the UV field and CRIR scaled up by a factor $\gamma = 30$, under the assumption that they are both produced (directly or indirectly) by nearby massive stars. This enhancement drastically changes the thermodynamics of the gas, with a corresponding impact on the physical evolution of the cloud, its chemical composition, and the resulting line emission properties \citep{clark2019,barnes2024,cusack2025}. We run the simulations until they have formed $\sim 100 \msun$ of sinks, which occurs after $5.53 \myr$ for $\gamma = 1$ and $7.31 \myr$ for $\gamma = 30$; simulation parameters are listed in Table \ref{tab:arepo}.

\subsection{Chemical modelling}

From each simulation, we randomly select $10^5$ tracer particles within $16.2 \pc$ of the centre of the computational domain at the simulation endpoint, chosen to uniformly span densities between $10-10^6 \pcc$; this density range encompasses the vast majority of the mass. The tracer particle histories are the basis for chemical modelling using the NEATH framework \citep{priestley2023b}, in which a modified version of the {\sc uclchem} code \citep{holdship2017} is used to evolve a reaction network based on UMIST12 \citep{mcelroy2013}. Elemental abundances are listed in Table \ref{tab:abun}, again taken from \citet{sembach2000}, with Mg, S and Si reduced by a factor of 100 for consistency with the {\sc arepo} `metal' abundance.

The NEATH chemical model from \citet{priestley2023b} was designed to reproduce the H$_2$ and CO abundances from the internal {\sc arepo} network under $\gamma = 1$ conditions; as these are the molecules most important for the thermal and dynamical evolution of the gas, we would not then expect any significant changes were we to rerun the MHD simulations with the full UMIST12 network (the computational expense of doing this is prohibitive in practice). Extending this approach to $\gamma = 30$ required substantial modifications to both the chemical model and the underlying mechanics of the {\sc uclchem} code, which are described in Appendix \ref{sec:neath}. We also found a better agreement between the NEATH and {\sc arepo} CO abundances by adopting the \citet{wakelam2010} rate for the C$^+$ + OH reaction, as used by \citet{gong2017}, rather than the default UMIST12 value; other differences in reaction rates between the two networks were not found to exert a significant influence on the results.

\subsection{Radiative transfer}

We perform line radiative transfer modelling using {\sc radmc3d} \citep{dullemond2012}, mapping the physical properties of the {\sc arepo} Voronoi mesh onto a cubic adaptively-refined grid such that each cell contains at most one Voronoi sampling point. Molecular abundances are then assigned to each cell from the nearest post-processed tracer particle. This approach accurately recovers the correct line emission properties, despite some loss of resolution compared to the original Voronoi mesh \citep{priestley2024}. Neighbouring cells can be assigned identical velocities, in which case the velocity divergence entering the large-velocity-gradient calculation of the line opacity is zero, resulting in unphysical behaviour. To avoid this, we set a maximum escape probability length scale of $16.2 \pc$, the half-width of the computational domain; the exact value of this parameter has a negligible effect on the results.

We produce position-position-velocity (PPV) cubes of the $^{12}$CO, HCN, HCO$^+$ and N$_2$H$^+$ $J=1-0$ transitions, with a spatial resolution\footnote{Note that these are the resolutions of the output PPV cubes, not those of the radiative transfer modelling. We use the {\sc radmc3d} recursive subpixelling and Doppler catching routines to avoid artifacts caused by resolution mismatches between the output cube and input grid of physical properties.} of $0.16 \pc$ and a velocity channel width of $0.034 \kms$. Collisional rates are taken from the LAMDA database \citep{schoier2005} for the partners listed in Table \ref{tab:lamda}. To reduce computational cost, we assume pure rotational spectra for HCN and N$_2$H$^+$, neglecting their hyperfine structure. These lines are typically optically thick, so this approach will tend to underestimate the integrated line intensity by a factor of a few, as the individual hyperfine components saturate more slowly than the single composite line \citep{priestley2023a}. The qualitative behaviour, however, is largely unchanged, as this is primarily governed by the physical-chemical structure of the simulation, and not by radiative transfer effects.

\begin{figure*}
  \centering
  \includegraphics[width=\columnwidth]{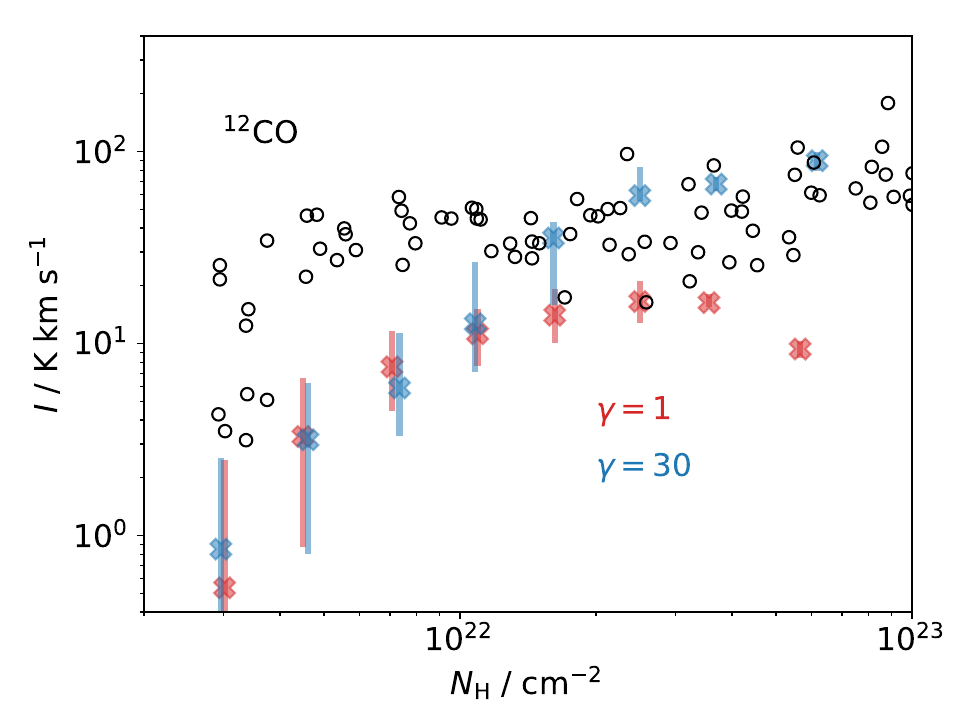}
  \includegraphics[width=\columnwidth]{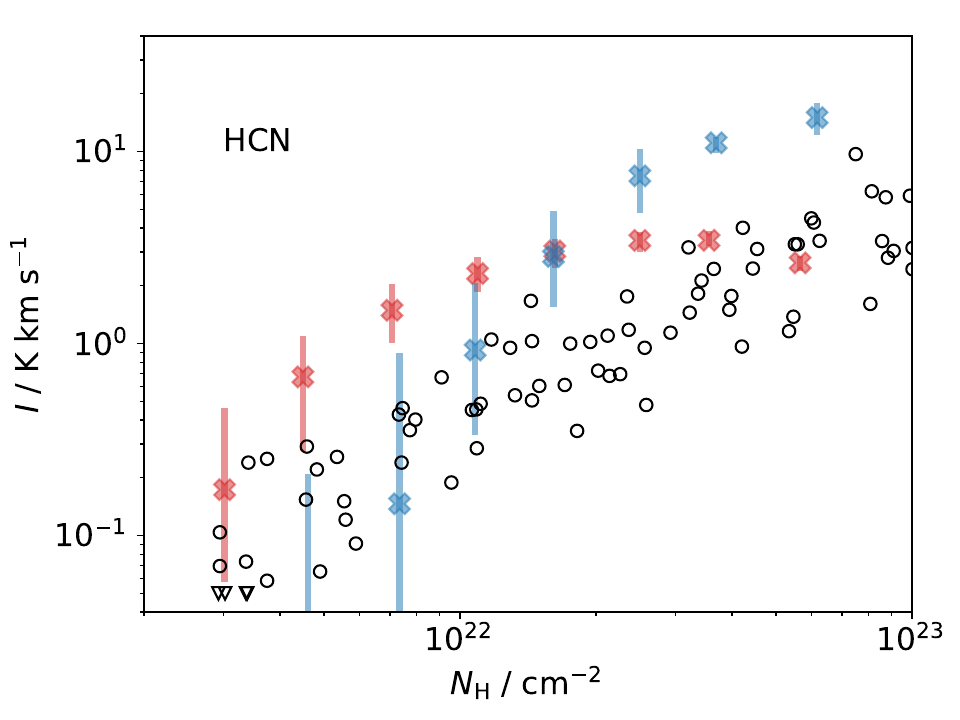}\\
  \includegraphics[width=\columnwidth]{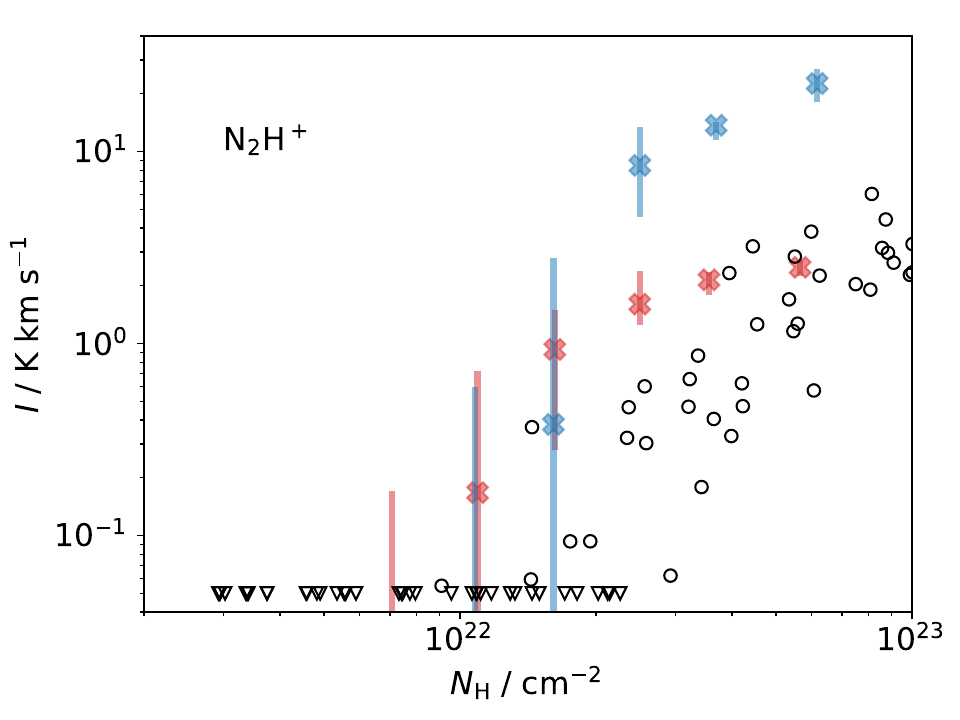}
  \includegraphics[width=\columnwidth]{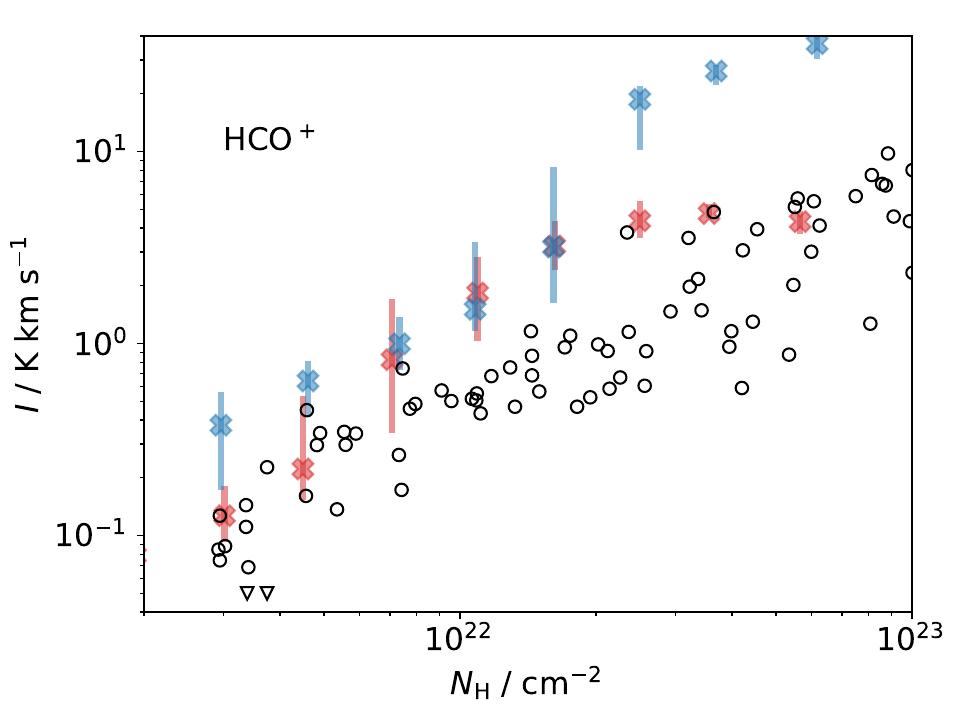}
  \caption{Integrated line intensities {of the $J=1-0$ transitions} versus column density for the $\gamma = 1$ (red) and $30$ (blue) simulations. The median values are shown as crosses, with the bars indicating the 16th/84th percentiles. Black circles show Perseus molecular cloud data from \citet{tafalla2021}, triangles indicate non-detections (for an assumed detection threshold of $0.05 \kel \kms$).}
  \label{fig:alpha}
\end{figure*}

\begin{figure*}
  \centering
  \includegraphics[width=\columnwidth]{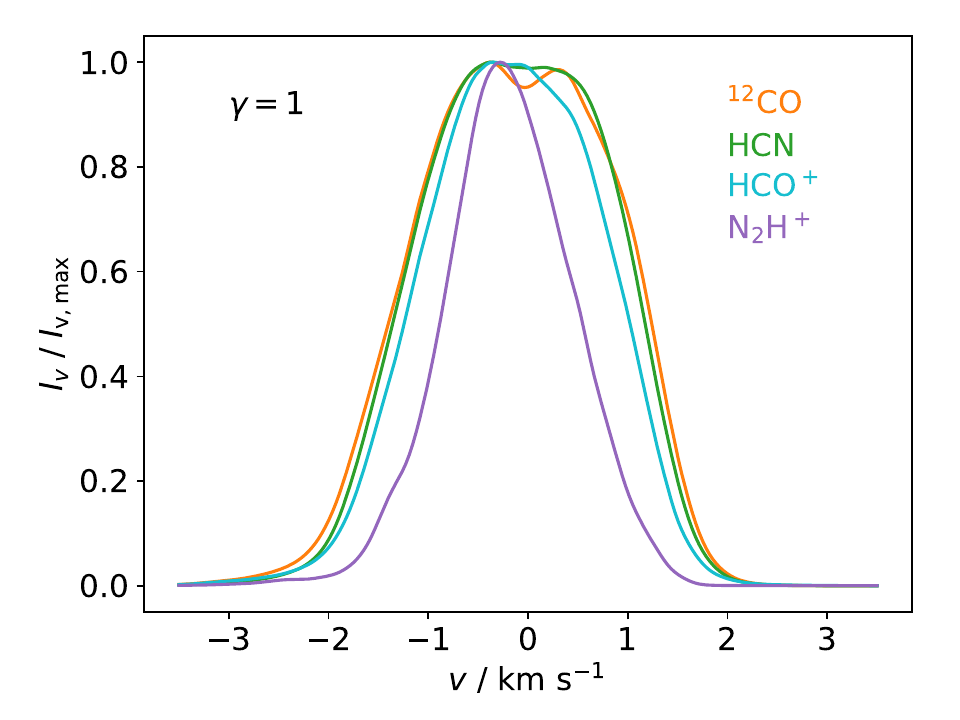}
  \includegraphics[width=\columnwidth]{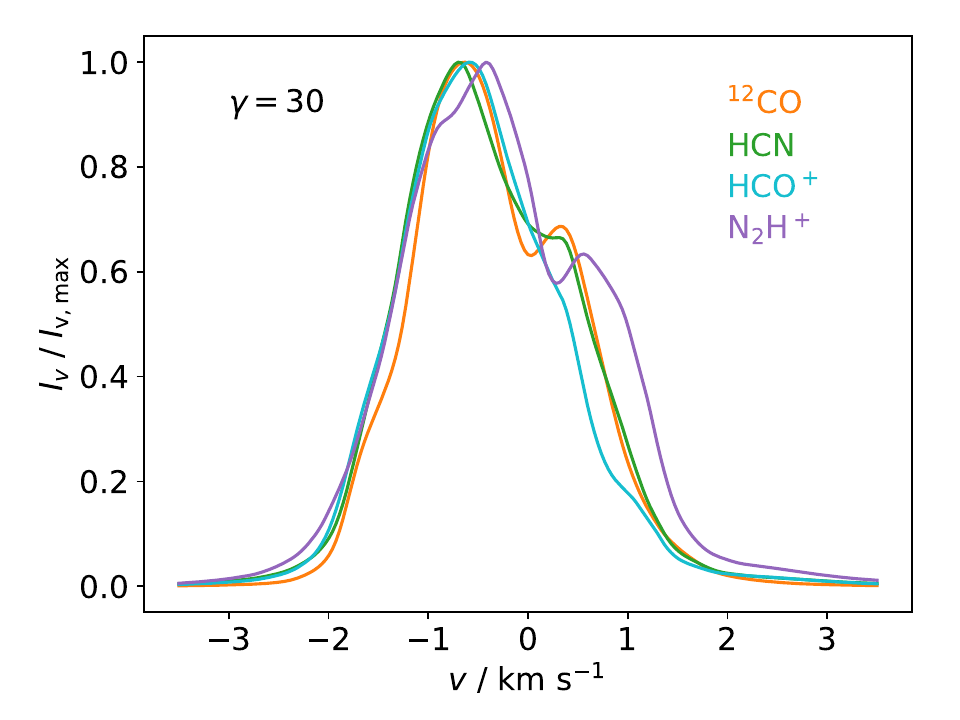}
  \caption{{Cloud-averaged line profiles of the $J=1-0$ transitions for the $\gamma = 1$ (left) and $30$ (right) simulations, normalised to the peak intensity of each line.}}
  \label{fig:profile}
\end{figure*}

\begin{figure*}
  \centering
  \includegraphics[width=\columnwidth]{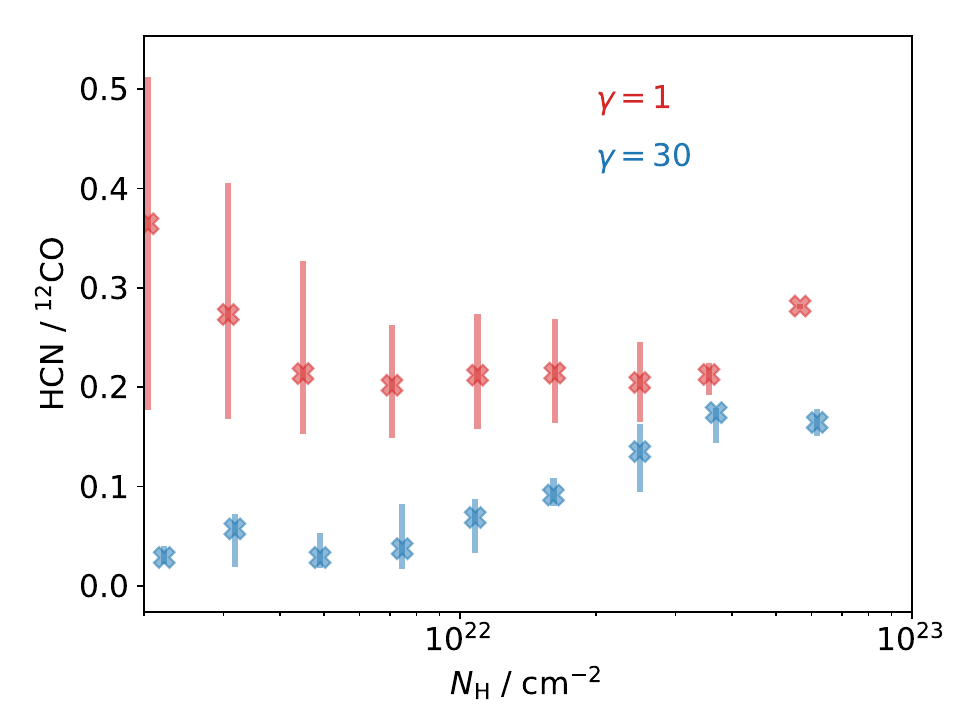}
  \includegraphics[width=\columnwidth]{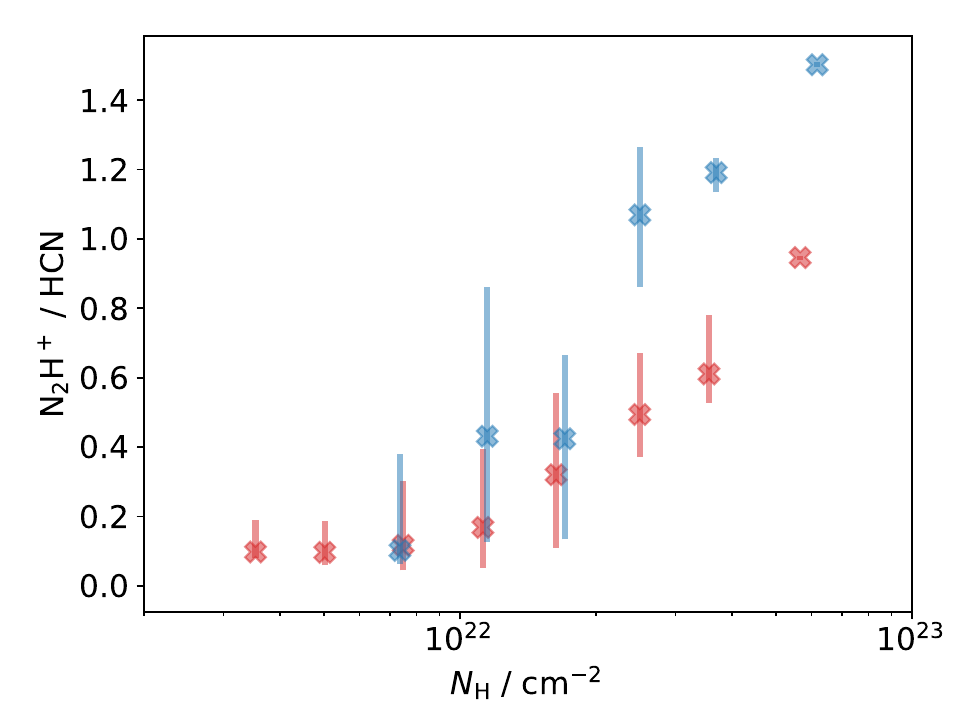}
  \caption{Line intensity ratios {of the $J=1-0$ transitions} versus column density for the $\gamma = 1$ (red) and $30$ (blue) simulations, ignoring pixels with intensities below $0.05 \kel \kms$. The median values are shown as crosses, with the bars indicating the 16th/84th percentiles.}
  \label{fig:ratio}
\end{figure*}

\begin{figure*}
  \centering
  \includegraphics[width=\columnwidth]{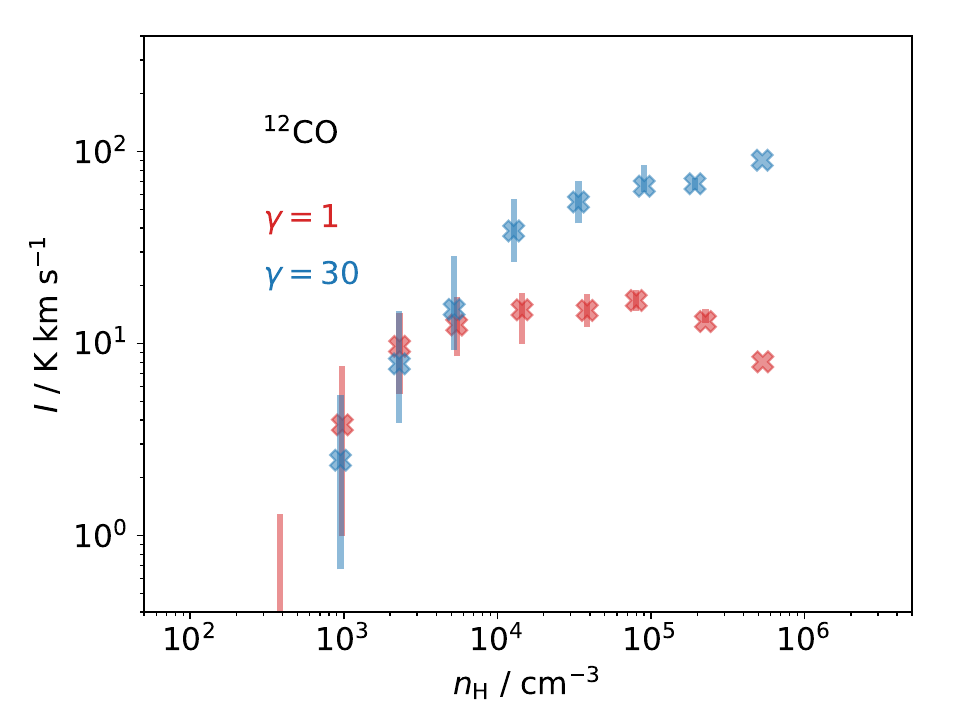}
  \includegraphics[width=\columnwidth]{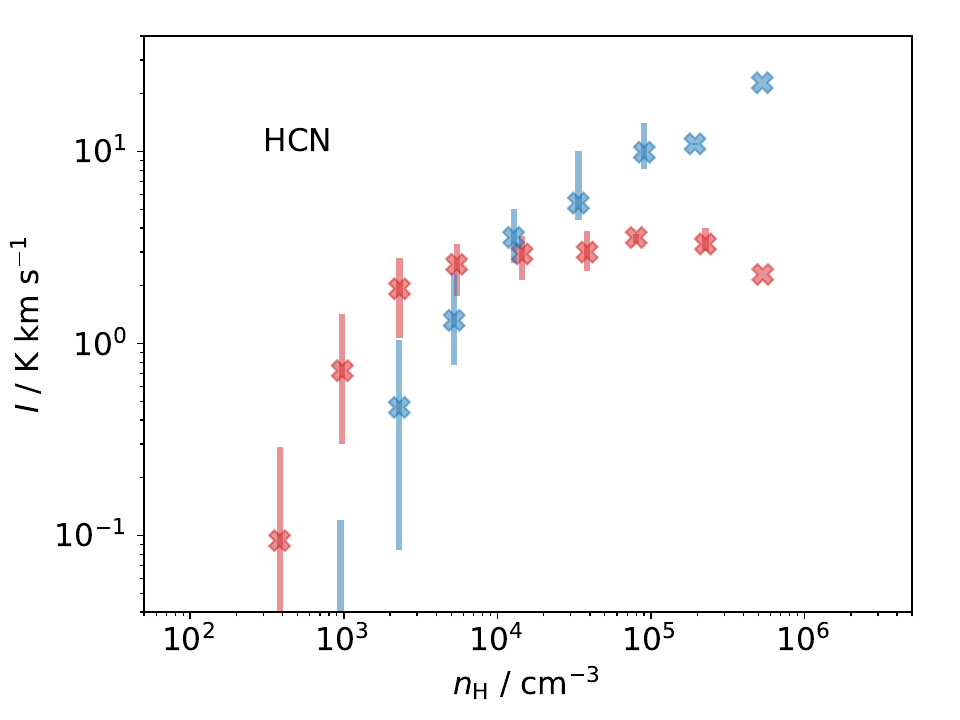}\\
  \includegraphics[width=\columnwidth]{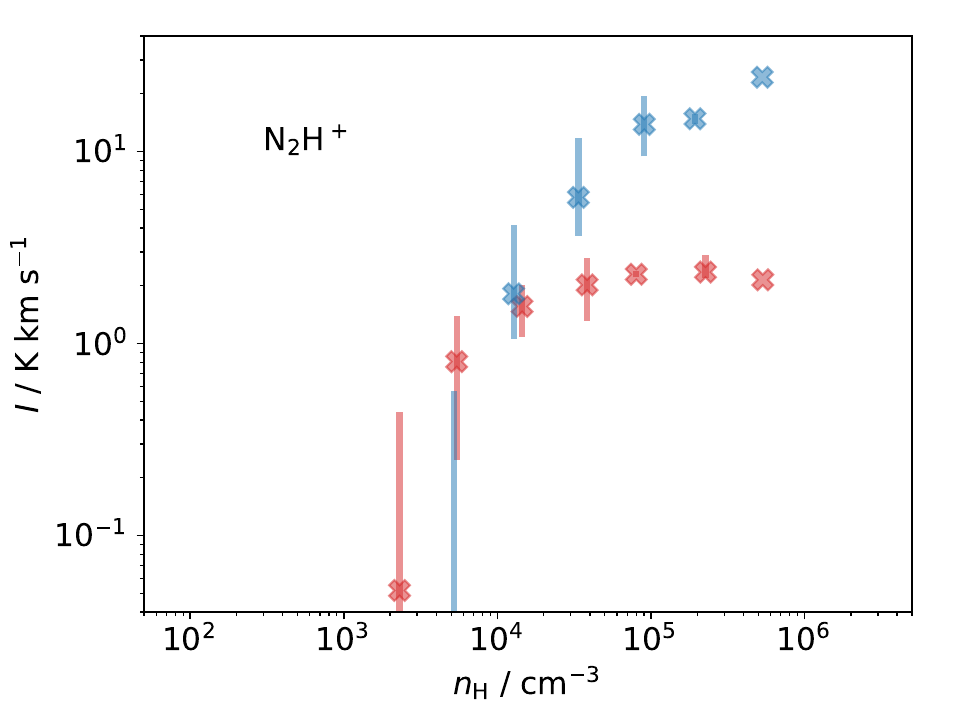}
  \includegraphics[width=\columnwidth]{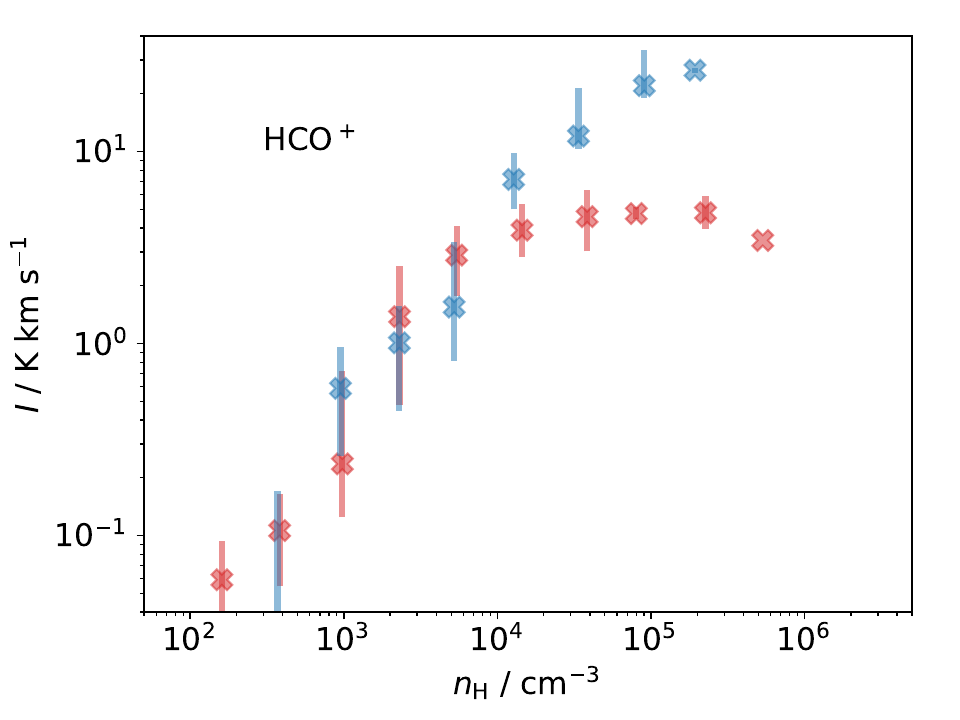}
  \caption{Integrated line intensities {of the $J=1-0$ transitions} versus peak line-of-sight volume density for the $\gamma = 1$ (red) and $30$ (blue) simulations. The median values are shown as crosses, with the bars indicating the 16th/84th percentiles.}
  \label{fig:peak}
\end{figure*}

\begin{figure*}
  \centering
  \includegraphics[width=\textwidth]{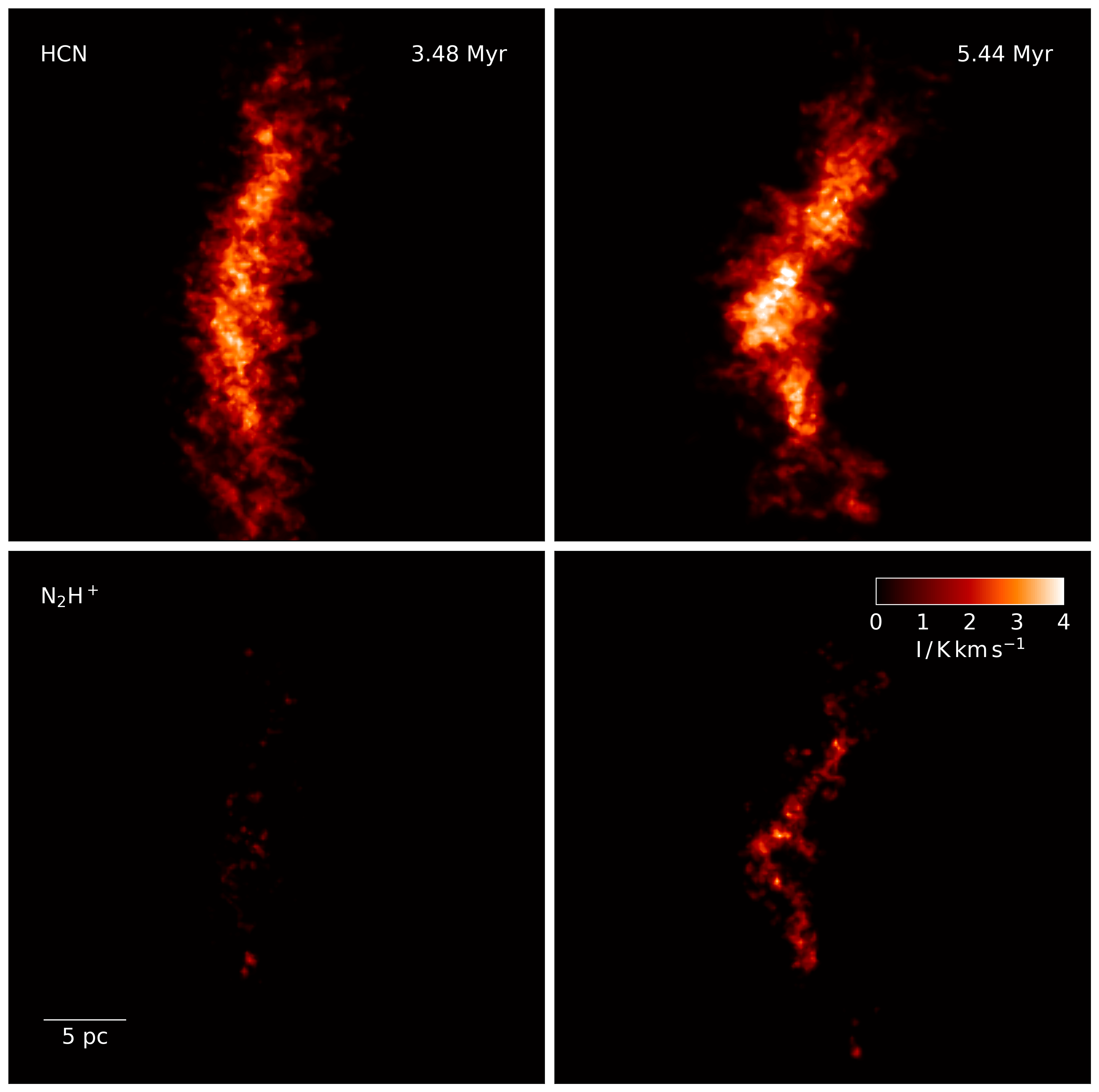}
  \caption{Integrated line intensity maps of HCN (top) and N$_2$H$^+$ (bottom) for the $\gamma = 1$ simulation after $3.48$ (left) and $5.44 \myr$ (right).}
  \label{fig:linesg1}
\end{figure*}

\begin{figure*}
  \centering
  \includegraphics[width=\columnwidth]{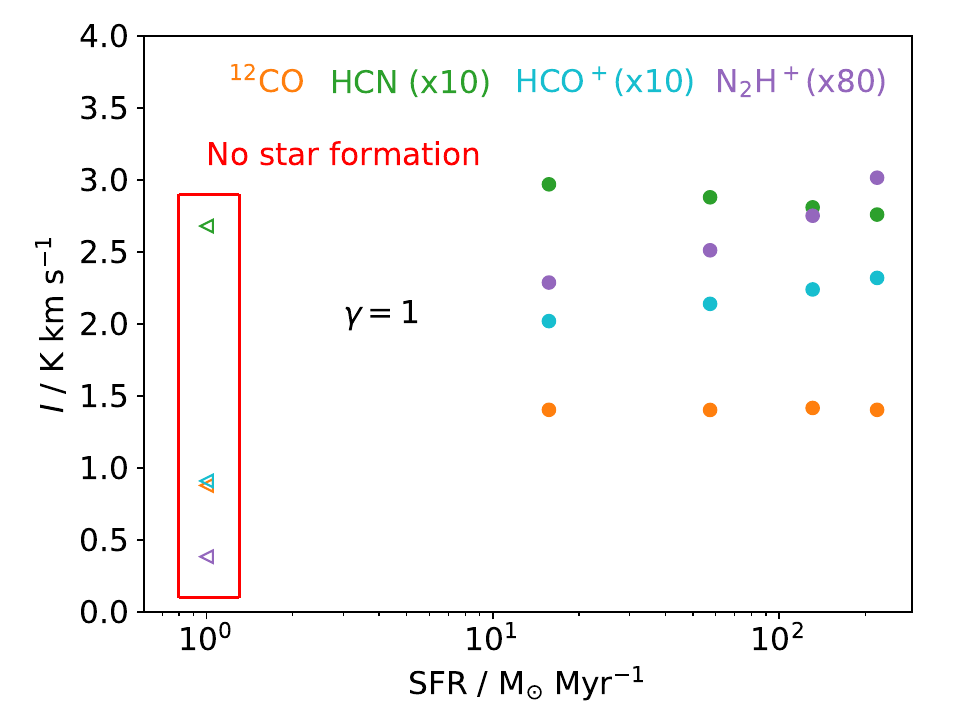}
  \includegraphics[width=\columnwidth]{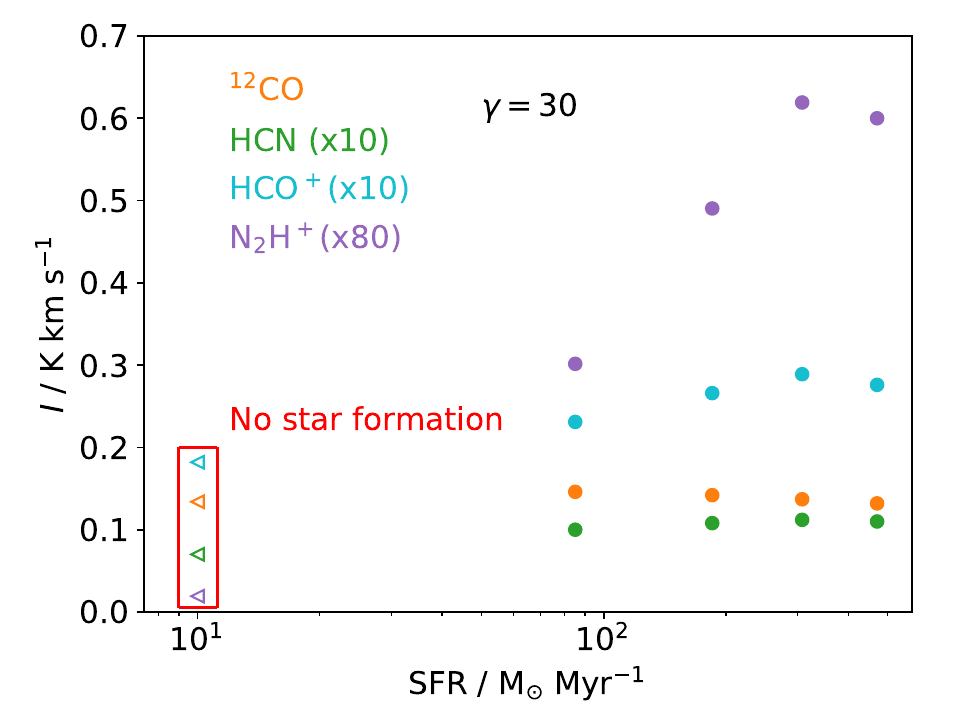}
  \caption{Cloud-averaged line intensities versus SFR for the $\gamma = 1$ (left) and $30$ (right) simulations, for $^{12}$CO (orange), HCN (green), HCO$^+$ (cyan) and N$_2$H$^+$ (purple). Intensities for lines other than $^{12}$CO have been scaled up by the amounts indicated in the figure to remain visible on the same axes.}
  \label{fig:linesfr}
\end{figure*}

\section{Results}

\subsection{Physical properties}

In Figure \ref{fig:facecol}, we show final column density maps for the two simulations, seen face-on to the collision (parallel to the collision velocity). The enhanced UV field and CRIR in the $\gamma = 30$ simulation results in dramatic changes to the physical cloud structure, going from a dispersed network of filaments to a much more centrally-condensed configuration reminiscent of Galactic hub-filament systems \citep{peretto2013,williams2018,anderson2021}. This is due to the higher thermal pressure of the cloud material acting to compress the locally-collapsing regions \citep{cusack2025}, while the higher temperatures result in an average sink mass a factor of two higher than the $\gamma = 1$ simulation, due to the increase in the typical Jeans mass (Table \ref{tab:arepo}).

Despite these very different morphologies, the two simulations have formed a comparable amount of dense gas: Figure \ref{fig:mass} shows the evolution of the mass accreted onto sinks, and the gas mass above density thresholds of $10^3$ and $10^4 \pcc$. The collision between clouds results in the rapid generation of moderate-density material within $1 \myr$, but with no associated star formation activity \citep{clark2012b,hunter2023}. In both cases, sink particles only begin to appear after the formation of a substantial ($\sim 100 \msun$) mass of gas above a density of $10^4 \pcc$. The assembly of this star formation reservoir is delayed by several $\myr$ in the $\gamma = 30$ simulation, but once in place, star formation occurs significantly more rapidly than for $\gamma = 1$ due to the more compact nature of the high-density regions \citep{cusack2025}.

That this apparent star formation threshold is unchanged between the two simulations is somewhat surprising: gas temperatures, shown in Figure \ref{fig:physprop}, are $2-3$ times higher in the $\gamma = 30$ simulation than for $\gamma = 1$, which one might expect would affect the characteristic density at which material becomes gravitationally bound. We note that while the gas temperature at a given density is much higher for $\gamma = 30$, the effective\footnote{The effective shielding column is the value of $N_{\rm H}$ that reproduces the angle-averaged $\exp \left(-2.5 A_{\rm V} \right)$ term responsible for attenuating UV photodissociation rates \citep{glover2012b}, {with $A_{\rm V} = 5.3 \times 10^{-22} \, \left(N_{\rm H} / {\rm cm^{-2}} \right) \, {\rm mag}$ \citep{bohlin1978}}.} shielding column density, also shown in Figure \ref{fig:physprop}, is almost unchanged; shielding from UV radiation has previously been identified as the key factor in determining whether material can collapse to form stars \citep{glover2012a,clark2014}. Given recent debate regarding the value \citep{offner2025} or existence \citep{moon2025} of any threshold density for star formation, we focus here on the relationship between the SFR and molecular line emission in the simulations, and leave a full investigation of star formation thresholds to future work.

\subsection{Chemical properties}

Molecular abundances versus density for the four species we consider here (CO, HCN, HCO$^+$ and N$_2$H$^+$) are shown in Figure \ref{fig:chem} at the respective endpoints of the $\gamma = 1$ and $30$ simulations. For $\gamma = 1$, the abundances of all four molecules increase with density up to around $10^4 \pcc$, before declining at higher densities due to freeze-out onto grain surfaces. The abundance increase at low densities is sharper for the two charged species, HCO$^+$ and N$_2$H$^+$, for which the primary cause is the decreasing ionisation fraction of the gas: free electrons undergo fast recombination reactions to destroy both species. For CO and HCN, the increasing abundance is instead due to the greater degree of shielding from photodissociation by UV photons.

The $\gamma = 30$ simulation has a significantly altered chemical structure, in addition to the physical differences discussed above. The enhanced UV field drastically reduces the abundances of CO and HCN at low densities via increased photodissociation reactions: {combined with the slightly lower effective shielding columns in this regime (Figure \ref{fig:physprop}), typical photodissociation rates are a factor $\gtrsim 50$ higher than for $\gamma = 1$ at a density of $100 \pcc$, and only a factor of a few lower than the baseline unshielded rate.} HCO$^+$ and N$_2$H$^+$ are less affected, because while the density of free electrons is higher, the subsequent increase in dissociative recombination reactions is partially compensated for by the increase in ionisations of H$_2$ {and other molecules}, the rate-limiting step for the formation of both species. At densities of $10^4 \pcc$ and above, the exponential dependence of photoreactions on the shielding column makes the enhanced radiation field irrelevant to the chemistry, {suppressed by factors of 100 or more for typical shielding columns in this regime} \citep{clark2013}. Differences here are driven mostly by the increase in cosmic ray-induced desorption reactions \citep{roberts2007}, which counteracts freeze-out and results in near-constant abundances for all species.

The sharp increase in abundance up to a density of $10^4 \pcc$ for all species in the $\gamma = 30$ simulation, particularly HCN, might suggest that these species would become better tracers of dense gas in these extreme environments. In W49, a massive star forming region with a presumably much higher ambient UV field than the Solar neighbourhood, \citet{barnes2020} found that HCN traces the same material as N$_2$H$^+$, rather than predominantly tracing more diffuse sightlines as in local molecular clouds \citep{kauffmann2017}. Conversely, N$_2$H$^+$ emission is found to be widespread in the Central Molecular Zone of the Milky Way \citep{santamaria2021}. This is undoubtedly an extreme environment \citep{henshaw2023}, but with relatively little star formation given its total mass of dense gas \citep{barnes2017}, suggesting that under these conditions N$_2$H$^+$ may trace the entire reservoir of molecular material rather than solely the dense, star-forming component \citep{barnes2024}.

However, an important consideration in a particular molecule's tendency to trace high-density gas is its abundance at {\it low} density \citep{priestley2023c}. At a density of $10^3 \pcc$, the HCN abundance has reached $\sim 10^{-10}$ in the $\gamma = 30$ simulation, easily sufficient to produce bright line emission, whereas the N$_2$H$^+$ abundance is four orders of magnitude lower and undetectable by any reasonable standard. A similar situation holds for HCO$^+$, where the key difference to N$_2$H$^+$ is that HCO$^+$ {forms from CO and is not destroyed by reactions with it} (proton transfer between CO and HCO$^+$ leaves the chemical state of the gas unchanged). {The simultaneous fall in ionisation fraction and rise in CO abundance as the density increases from $10^2$ to $10^4 \pcc$ both promote increasing HCO$^+$ abundances, whereas the latter effect largely cancels out the former for N$_2$H$^+$ until the CO abundance stops rising.} The chemical fragility of N$_2$H$^+$, easily destroyed by both free electrons and by gas-phase CO, restricts it to undetectable levels in the moderate-density gas that makes up most of the cloud mass (Figure \ref{fig:mass}), only becoming abundant in the high-density material associated directly with star formation. This property is not shared by either HCN or HCO$^+$, with consequences for their line emission which we explore in the following section.

{The molecular abundances shown in Figure \ref{fig:chem} are generally not in chemical equilibrium: most of the dispersion at fixed density is due to the different evolutionary histories of the tracer particles, as the variations in temperature and shielding column are relatively small (Figure \ref{fig:physprop}). However, chemical reaction timescales are still short enough for the molecular abundances to respond quickly (on timescales of $100 \kyr$ or less) to changes in the physical conditions \citep{holdship2022,priestley2023b}. Once conditions favourable for N$_2$H$^+$ formation arise, it quickly reaches detectable levels in the gas phase, and if conditions then become unfavourable again it is destroyed equally rapidly \citep{priestley2023c}. As such, it is rare to find either star-forming gas with a low N$_2$H$^+$ abundance, or non-star-forming gas with significant quantities of N$_2$H$^+$, despite the potential for non-equilibrium chemistry to disrupt the association between the instantaneous physical conditions and the molecular abundances of cloud material.}

\subsection{Line emission}

Figure \ref{fig:alpha} compares the integrated line intensities as a function of line-of-sight column density to observations of the Perseus molecular cloud from \citet{tafalla2021}; other local molecular clouds have comparable line emission properties to Perseus \citep{pety2017,tafalla2023}, making these data representative of a `normal' star-forming cloud. The $\gamma = 30$ simulation produces stronger line emission along high-column sightlines than the $\gamma = 1$ case, due to the enhanced CRIR raising the gas temperature. At lower columns, where {the $J=1-0$} lines are subthermally excited and less sensitive\footnote{{At a density of $10^3 \pcc$, {\sc radex} calculations \citep{vandertak2007} suggest that the increase in temperature from $20$ to $100 \kel$ (as in Figure \ref{fig:physprop}) only results in a $10-20\%$ increase in excitation temperature and intensity for the lines in question.}} to the temperature \citep{bemis2024}, the intensities are comparable for $^{12}$CO, N$_2$H$^+$, and HCO$^+$. The exception is HCN, where enhanced photodissociation significantly reduces its abundance in low-density gas for $\gamma = 30$ (Figure \ref{fig:chem}), and thus its line emission. The other species are either primarily destroyed by recombination (N$_2$H$^+$, HCO$^+$), or efficiently self-shield from UV photons at moderate densities (CO), so are less affected.

Both simulations underestimate the Perseus $^{12}$CO line intensity at low columns by a significant factor. As the line is optically thick even at these column densities, the discrepancy suggests that the molecular gas in our simulations covers a significantly narrower velocity range than it does in Perseus. {Figure \ref{fig:profile} shows the cloud-averaged line profiles of the $J=1-0$ transitions for the four molecules in both simulations. The $^{12}$CO line widths of $3-4 \kms$ are significantly narrower than the $6-8 \kms$ seen in Perseus by \citet{tafalla2021}. In the $\gamma = 1$ simulation, the HCN and HCO$^+$ lines also have widths of $3-4 \kms$, while N$_2$H$^+$ is somewhat narrower, which is in reasonable agreement with the Perseus data. The N$_2$H$^+$ linewidth is instead comparable to or even greater than that of HCN and HCO$^+$ in the $\gamma = 30$ simulation; the higher densities and enhanced N$_2$H$^+$ abundances result in its emission tracing the bulk kinematics of the cloud, rather than solely that of the densest regions as in the $\gamma = 1$ case \citep{barnes2024}.}

{The relatively narrow $^{12}$CO line widths in our simulations are} not entirely surprising, as we consider colliding clouds of initially-atomic gas, which form a compressed layer of molecular material with a substantially-reduced velocity dispersion from the initial turbulent field \citep{priestley2023a}. Real molecular clouds are unlikely to conform to this idealised setup, which will by construction underproduce the amount of extended, moderate-density (and therefore CO-emitting) material along low-column sightlines. In future work, we will explore whether less-idealised, kpc-scale simulations of the turbulent ISM result in more `normal' distributions of CO line emission (Clark et al. in prep.).

The range of `dense gas tracer' line intensities in the $\gamma = 1$ simulation is comparable to the Perseus data, whereas it systematically underproduces $^{12}$CO emission at all columns. Conversely, the $\gamma = 30$ simulation is in good agreement with the Perseus data for $^{12}$CO at high columns, but overproduces emission from the other three lines. This suggests that the temperature in regions being probed by the dense gas tracers is higher in $\gamma = 30$ than in Perseus, but the temperature in the lower-density regions reponsible for the bulk of the $^{12}$CO emission is comparable, with the reverse being the case for $\gamma = 1$ (too cold at low density, comparable at high density). As the dominant heating mechanisms at high and low density are, respectively, the CRIR and the background UV field \citep{cusack2025}, the Perseus cloud - and, by extension, other nearby star-forming regions \citep{tafalla2023} - might be best described by a simulation with a standard CRIR and an enhanced UV field. We discuss this possibility further in Section \ref{sec:radfield}.

The increased radiation field and subsequently-reduced HCN abundance in the $\gamma = 30$ simulation result in its line intensity behaving more similarly to N$_2$H$^+$, rising sharply towards the \citet{lada2010} star formation threshold of $10^{22} \pcs$. The HCN/$^{12}$CO and N$_2$H$^+$/HCN line ratios, both commonly used observationally to trace the fraction of dense gas, are shown versus column density in Figure \ref{fig:ratio}. Under standard Solar neighbourhood conditions, the HCN/$^{12}$CO ratio is insensitive to column density \citep{priestley2024}, but for the $\gamma = 30$ simulation it systematically increases beyond columns of $\sim 10^{22} \pcs$, as does the N$_2$H$^+$/HCN ratio in both simulations. This is in agreement with observations of the W49 high-mass star-forming region \citep{barnes2020}, with its presumably-higher ambient UV field; under these conditions, HCN traces a similar column density regime to N$_2$H$^+$.

Although the above might suggest that HCN also traces the star-forming material under extreme conditions, the SFR in our simulations correlates with the mass of high {\it volume} density gas, not column. Figure \ref{fig:peak} shows integrated line intensities versus the peak volume density along the line of sight for each pixel. The $^{12}$CO, HCN and HCO$^+$ lines all produce non-negligible (i.e. above $0.1 \kel \kms$) intensities for peak line-of-sight densities of $\sim 10^3 \pcc$, whereas the N$_2$H$^+$ line remains undetectable until the peak volume density approaches the $10^4 \pcc$ threshold related to the onset of star formation in the simulations. This holds even for the $\gamma = 30$ simulation: despite being a better tracer of regions with high column density, HCN is still not specifically tracing the star-forming component of the clouds.

Figure \ref{fig:linesg1} shows edge-on line emission maps from the $\gamma = 1$ simulation at two different snapshots, spanning the onset of star formation so that the instantaneous SFR\footnote{We define the SFR using the change in sink mass between {\sc arepo} snapshots; full details are given in Appendix \ref{sec:data}.} increases from zero to a few $100 \msun \myr^{-1}$. Despite this increase in the SFR, the cloud's HCN emission is little-changed, with some central regions becoming modestly brighter. The increase in N$_2$H$^+$ emission is much more significant, going from a few scattered, low-brightness clumps in the pre-star formation cloud to the entire high-density spine of the cloud being visible once star formation is established. Similar results are obtained for the $\gamma = 30$ simulation, despite its very different physical and chemical structure.

We demonstrate the relationship between line emission and star formation quantitatively in Figure \ref{fig:linesfr}, where the average line intensities\footnote{We note that these values are very low compared to those seen in extragalactic observations, particularly for the N$_2$H$^+$ line; this is discussed in Section \ref{sec:beam}.} within the central $32.4 \pc$ are plotted versus the SFR at multiple timesteps for both simulations (these data are described and provided in tabular form in Appendix \ref{sec:data}). The HCO$^+$ and N$_2$H$^+$ lines show positive trends with SFR, albeit weak for HCO$^+$, but the most notable feature is the jump in N$_2$H$^+$ emission associated with the onset of star formation, with the integrated line intensity being effectively zero prior to the formation of the first sink particles. The $^{12}$CO and HCN lines show no positive trend with SFR; HCN emission even declines slightly as the SFR increases from $10$ to $100 \msun \myr^{-1}$ in the $\gamma = 1$ simulation. As anticipated by observational studies of local molecular clouds, there is no Gao-Solomon relationship {on these cloud scales of $30-50 \pc$ and below}: the HCN $J=1-0$ line is not a tracer of the star-forming gas.

\section{Discussion}

\begin{figure*}
  \centering
  \includegraphics[width=\columnwidth]{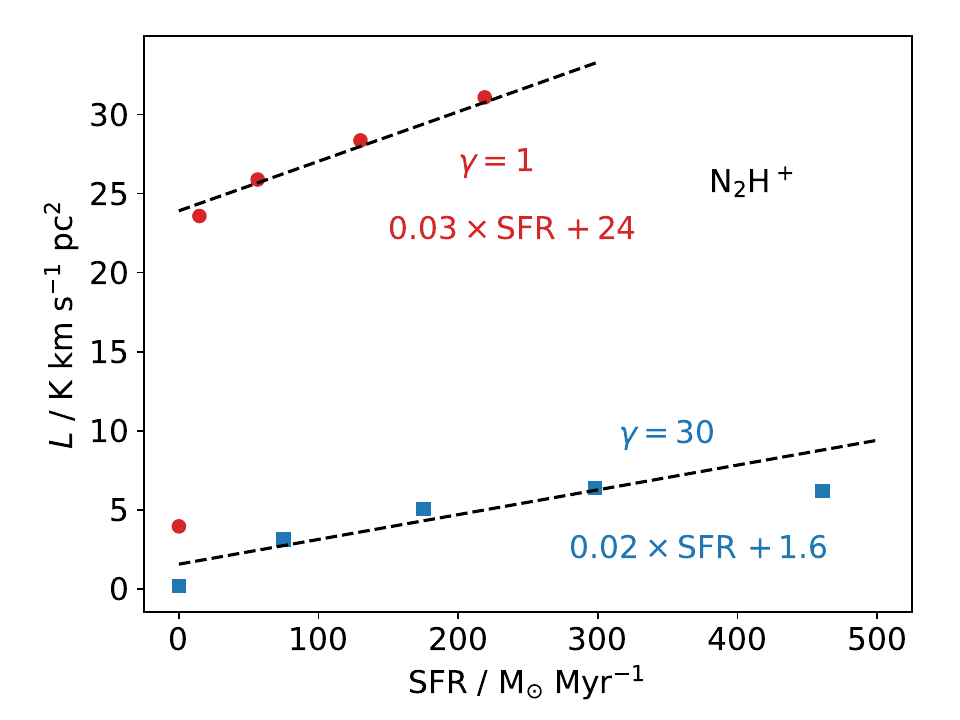}
  \includegraphics[width=\columnwidth]{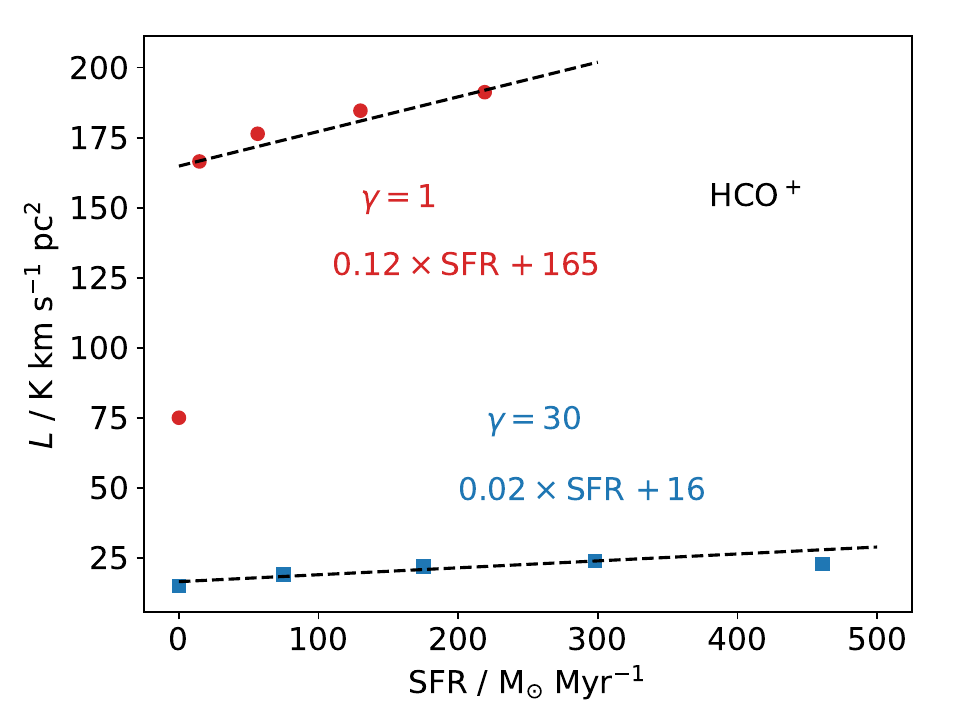}
  \caption{N$_2$H$^+$ (left) and HCO$^+$ (right) $J=1-0$ line luminosities versus SFR for the $\gamma = 1$ (red) and $30$ (blue) simulations. Dashed lines indicate the approximate linear relationship for the points with non-zero SFR.}
  \label{fig:linear}
\end{figure*}

\begin{figure*}
  \centering
  \includegraphics[width=\columnwidth]{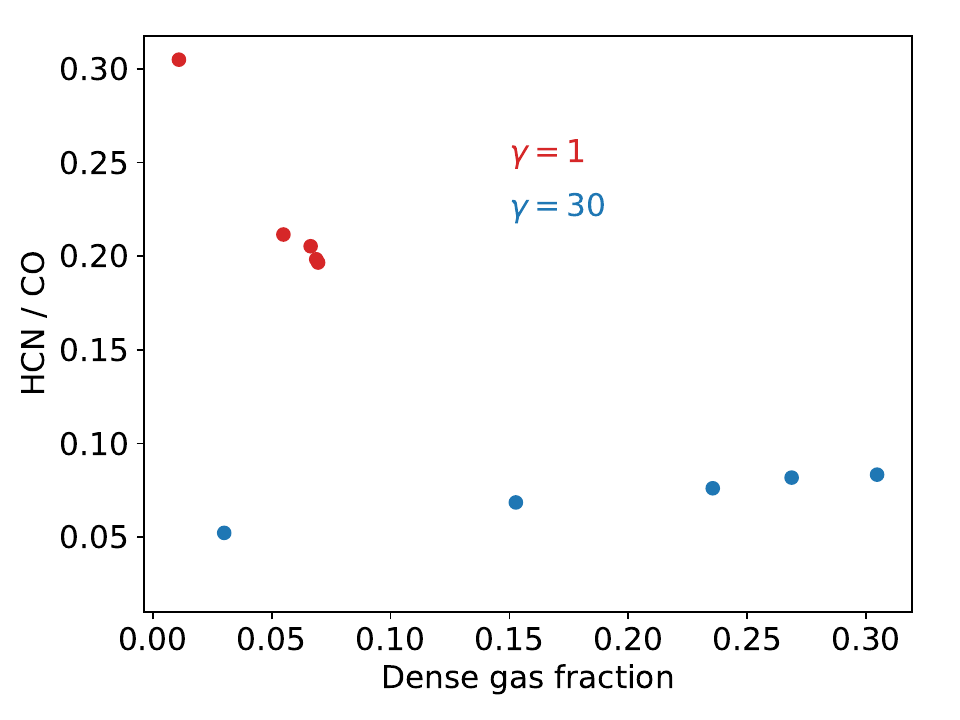}
  \includegraphics[width=\columnwidth]{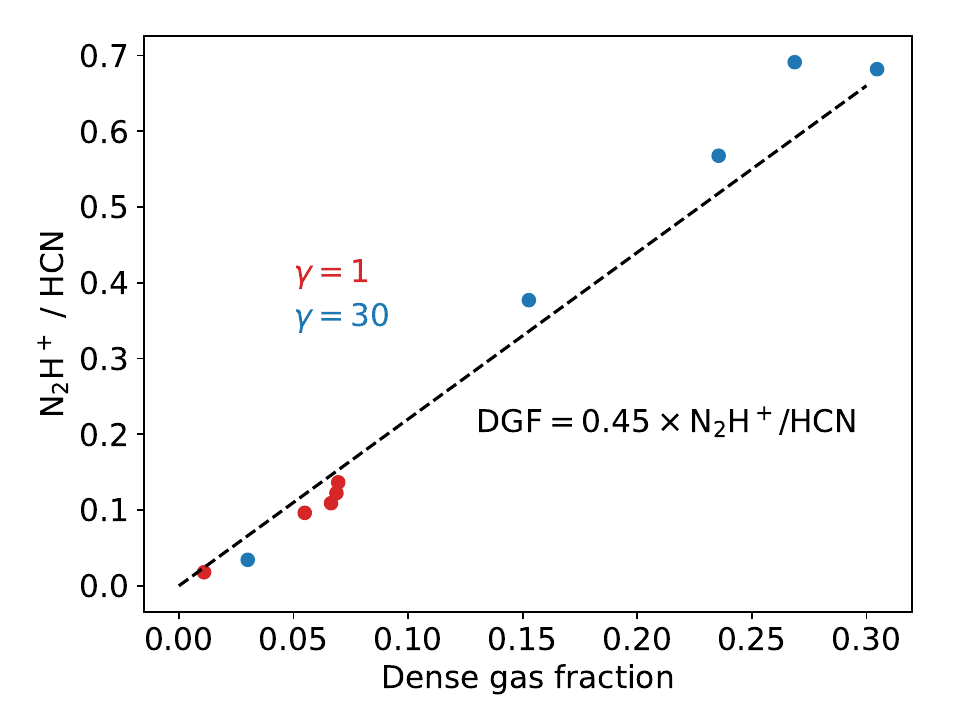}
  \caption{Line intensity ratios of HCN/$^{12}$CO (left) and N$_2$H$^+$/HCN (right) versus the dense gas fraction, for the $\gamma = 1$ (red) and $30$ (blue) simulations. The dense gas fraction is defined as the ratio of the mass in the simulation above a density of $10^4 \pcc$, linked with star formation activity, to the mass above $10^3 \pcc$, which is fully-molecular in both simulations. The dashed black line in the right panel indicates the approximate linear relationship between the dense gas fraction and the N$_2$H$^+$/HCN line ratio.}
  \label{fig:dense}
\end{figure*}

\begin{figure}
  \centering
  \includegraphics[width=\columnwidth]{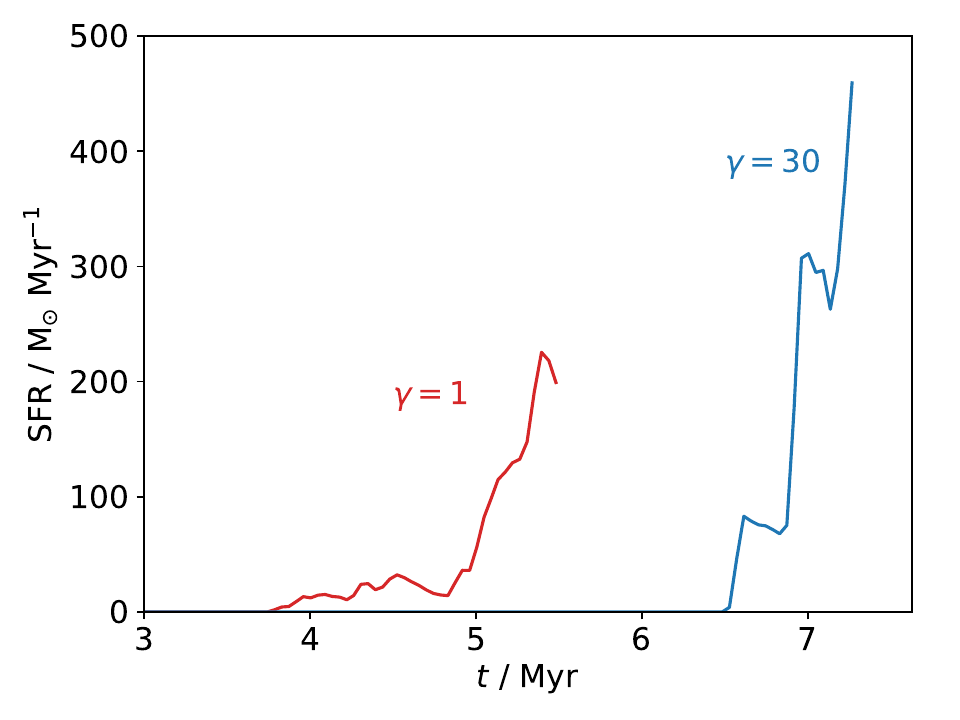}
  \caption{Instantaneous SFR (the difference in accreted mass between timesteps) versus time for the $\gamma = 1$ (red) and $30$ (blue) simulations.}
  \label{fig:sfr}
\end{figure}

\begin{figure*}
  \centering
  \includegraphics[width=\columnwidth]{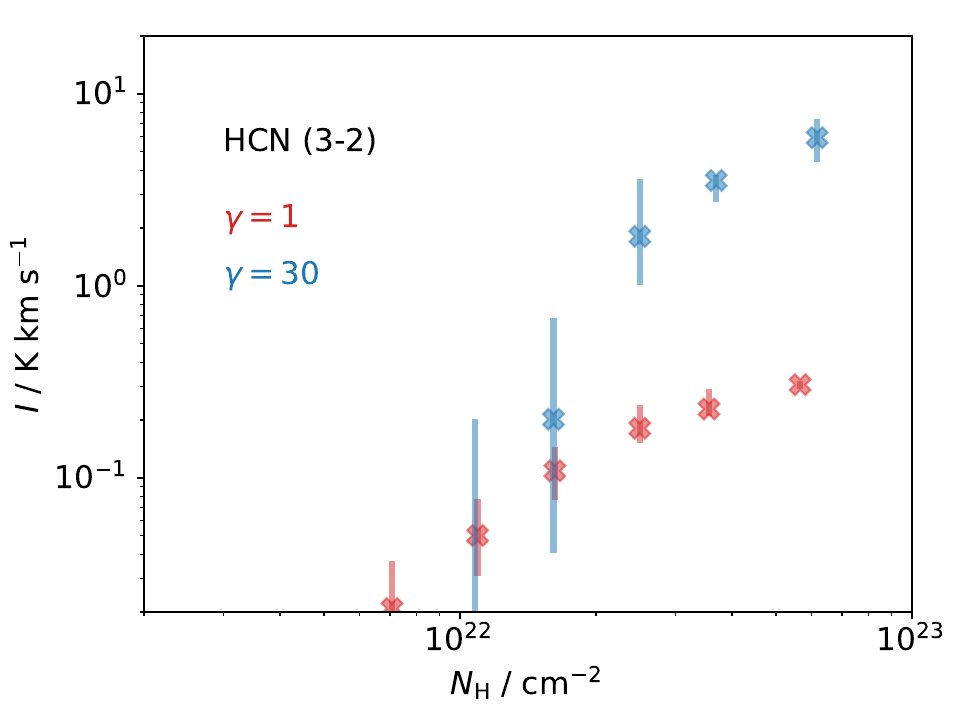}
  \includegraphics[width=\columnwidth]{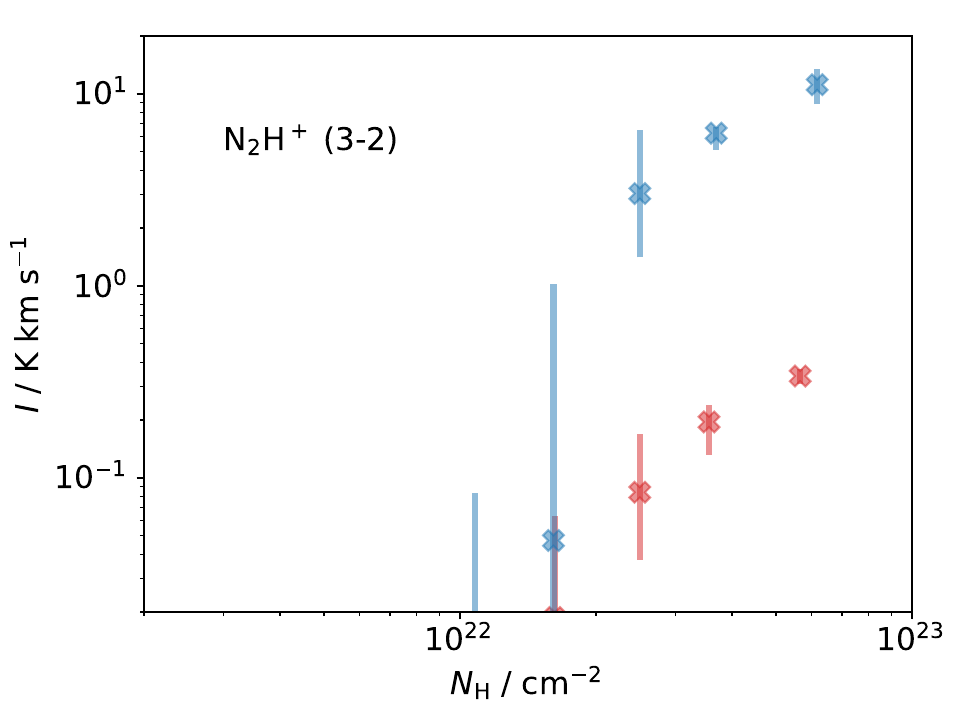}
  \caption{{Integrated line intensities of the HCN and N$_2$H$^+$ $J=3-2$ transitions versus column density for the $\gamma = 1$ (red) and $30$ (blue) simulations. The median values are shown as crosses, with the bars indicating the 16th/84th percentiles.}}
  \label{fig:alpha32}
\end{figure*}

\begin{figure}
  \centering
  \includegraphics[width=\columnwidth]{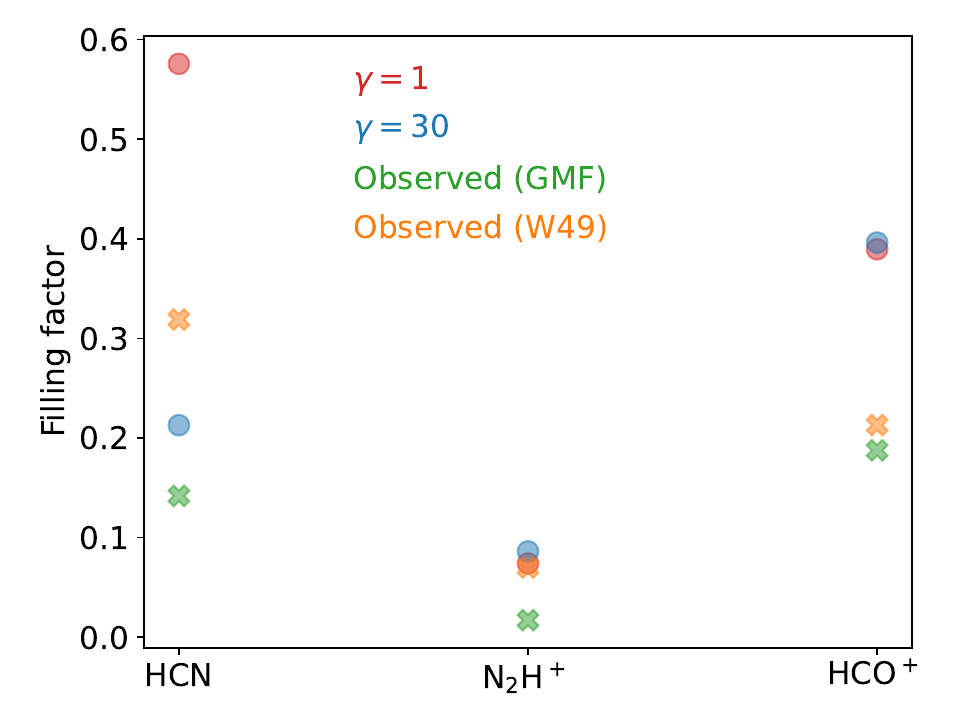}
  \caption{Line emission filling factors relative to CO for the $\gamma = 1$ (red) and $30$ (blue) simulations, with an assumed detection threshold of $1 \kel \kms$. Green crosses show average values for GMFs from \citet{feher2025}, orange crosses show W49 values from \citet{barnes2020}.}
  \label{fig:fillfac}
\end{figure}

\subsection{The origin of the Gao-Solomon relationship}

The simulations presented here reinforce what was already apparent from observational studies of nearby molecular clouds \citep{lada2010,evans2020}: HCN emission does not trace the reservoir of high-density gas directly associated with star formation. Extrapolating to the scales of entire galaxies, one would not expect anything resembling the Gao-Solomon relationship to exist. The fact that it does exist, across a huge range of both SFR and HCN luminosity \citep{neumann2025}, requires explanation.

We argue that the most likely reason for its existence is that the dense gas fraction (DGF) of molecular clouds is constant when measured over sufficiently large scales (or equivalently, sample sizes). While the DGF of local clouds does vary significantly from cloud to cloud \citep{lada2010}, these variations are driven by relatively minor changes in the power-law tails of their column density distributions \citep{lombardi2015}. It seems plausible that when considering all clouds within a kpc-sized region, as in extragalactic observations, the overall DGF of the cloud ensemble does not change significantly from region to region (or from galaxy to galaxy). \citet{urquhart2021} find that their sample of $936$ Milky Way clouds has a log-normal distribution of DGFs with a width of $0.45$ dex, narrow enough to present a clearly-defined mean value. With HCN emission effectively tracing the total mass of molecular gas \citep{tafalla2021,tafalla2023}, the Gao-Solomon relationship arises because the fraction of this gas actually involved in forming stars is universal.

This view is additionally supported by the relatively constant value of the N$_2$H$^+$/HCN line ratio found by \citet{jimenez2023}, measured over a wide range of size scales. We show in Section \ref{sec:dense} that this ratio is very strongly correlated with the DGF in the simulations, so the constant line ratio would appear to imply a similarly-constant DGF. On the smaller scales of individual clouds, this constant DGF/line ratio should break down, and it does: Orion A has a higher DGF, a higher SFR, and a higher N$_2$H$^+$/HCN ratio than Orion B, with both clouds having comparable total masses \citep{lada2010,jimenez2023}. We argue that the Gao-Solomon relationship is an emergent property of ensembles of molecular clouds, not a result of HCN tracing star-forming gas directly.

\subsection{Effects of stellar feedback}

\citet{clark2015} have suggested that stellar feedback may explain the Kennicutt-Schmidt relationship, at least in part: high star formation rates lead to higher gas temperatures due to the increase in UV radiation, which then increases the CO emissivity per unit mass. If the mass of molecular gas is measured assuming a constant CO emissivity, galaxies with higher SFRs will appear to have higher molecular masses than otherwise-identical ones with lower SFRs, producing a Kennicutt-Schmidt-like relationship even in the absence of any real underlying relationship between the variables. Our $\gamma = 30$ simulation is designed to represent molecular clouds in active star-forming environments, but its HCN emissivity is only modestly enhanced over the $\gamma = 1$ case at high column density (Figure \ref{fig:alpha}), and severely reduced at lower columns, so it appears unlikely that the mechanism proposed in \citet{clark2015} can also explain the Gao-Solomon relationship. However, our simulations only include the effects of the external UV field: feedback from newly-formed stars within the simulated clouds is not modelled.

As stars form in the highest-density regions of molecular clouds, the effect of radiative feedback is initially\footnote{Later stages involve the disruption of the cloud and destruction of its molecular content, so cannot result in any increase in line emission.} to raise the surrounding gas and dust temperatures to a few times the ambient cloud values \citep{bate2009}. We approximate the effect of this early feedback by imposing a temperature floor in the radiative transfer modelling, representing the extreme case where the entire cloud has been heated up by embedded protostars. Even imposing a minimum temperature of $50 \kel$, the HCN luminosity of the $\gamma = 1$ simulation increases by less than $10\%$. Most of the emission comes from moderate-density gas \citep{jones2023}, which is already at temperatures of $30-40 \kel$ and where {the HCN $J=1-0$ line} is subthermally excited \citep{bemis2024}, so the increased temperature has a minimal effect on the line emission strength. This again appears insufficient to reproduce the observed linear SFR-HCN relationship. We will investigate the impact of radiation from newly-formed stars self-consistently, using the radiative transfer capabilities in {\sc arepo} \citep{peter2023}, in future work.

\subsection{Low simulated line intensities}
\label{sec:beam}

The cloud-averaged intensities of the lines presented in Figure \ref{fig:linesfr} are low compared to those found in $\kpc$-scale observations of external galaxies. In particular, the integrated N$_2$H$^+$ intensity of our simulated clouds would barely reach reach the detection limits of modern extragalactic surveys \citep[e.g.][]{stuber2025a}. This is not an issue of the simulations underestimating line emission from the cloud material; in fact, they tend to produce too much line emission from dense gas tracers per unit column, compared to nearby molecular clouds (Figure \ref{fig:alpha}). The issue is instead the filling factor of high-density material, which is low in both simulations.

For any column-intensity relationship which reproduces the trends seen in nearby molecular clouds, an extragalactic detection of the N$_2$H$^+$ line would require a large fraction of the telescope beam to be filled with high-column material, with a beam-averaged column density of $10^{22} \pcs$ or above: essentially the entire beam should be filled by infrared-dark clouds such as those studied by \citet{rigby2024}. Resolved studies of kpc-scale regions of the Milky Way disc find the filling factor of N$_2$H$^+$-bright material is below $10\%$ \citep{wang2020,feher2024,feher2025}, although this is much higher in the Central Molecular Zone \citep{santamaria2021}. Similarly, resolved studies of the ISM of five nearby disc galaxies with $\sim 10 \pc$ resolution \citep{faustino2024,faustino2025} find only one (M51) with any appreciable fraction of material above $10^{22} \pcs$ on these scales, which is undergoing a strong interaction with a companion galaxy; this is also one of the few extragalactic systems with resolved N$_2$H$^+$ detections \citep{stuber2023}. Situations where the N$_2$H$^+$ line can be detected on extragalactic scales likely represent extreme environments, which are not representative of `normal' star formation in the disc of the Milky Way and other galaxies.

\subsection{Does any line trace the SFR?}
\label{sec:dense}

While the $^{12}$CO and HCN lines show no correlation with the SFR in our simulations, there are some signs of a positive relationship for HCO$^+$ and N$_2$H$^+$. For the N$_2$H$^+$ line, we previously hypothesised a linear relationship between the line luminosity and the SFR \citep{priestley2023c}. The actual relationship in the simulations, shown in Figure \ref{fig:linear}, is not linear, primarily due to the discontinuous jump in luminosity related to the onset of star formation in the clouds. Ignoring the data points at an SFR of zero, the simulations do produce a tentatively linear relationship, albeit with a non-zero intercept and a different slope for the $\gamma = 1$ and $30$ models. HCO$^+$ behaves similarly, but with a much larger intercept, so that the increase in line emission due to the SFR is minimal as a fraction of the overall line brightness. While we do find that the N$_2$H$^+$ and HCO$^+$ lines, unlike HCN, trace the SFR, the weakness of the trends, their environmental dependence, and, for N$_2$H$^+$, the inherent faintness of the line all restrict their usefulness as observational tools.

The N$_2$H$^+$ slopes are steeper by a factor of about ten than the one we proposed in \citet{priestley2023c}, which predicted a purely-linear relationship with no intercept. This relied on two assumptions which are not realised in these simulations: the intensity-column relationship was taken from \citet{tafalla2021}, which unlike our simulations is linear above $10^{22} \pcs$; and all gas traced by N$_2$H$^+$ was assumed to be uniformly collapsing on a free-fall timescale at a density of $10^4 \pcc$, which is not the case in these simulations. In particular, one can see in Figure \ref{fig:mass} that there is generally a delay of almost a Myr before changes in the dense gas reservoir affect the mass being accreted by sinks. As such, we suggest that detectable N$_2$H$^+$ emission remains an excellent signpost of imminent or ongoing star formation, but not one which can be used to quantitatively assess the SFR.

What N$_2$H$^+$ {\it can} be used to trace is the amount of dense gas, specifically that above the $10^4 \pcc$ density threshold required for its formation. Figure \ref{fig:dense} shows the line ratios of HCN/$^{12}$CO and N$_2$H$^+$/HCN, integrated over the whole cloud, versus the DGF, defined here as the ratio of the cloud mass above $10^4 \pcc$ to that above $10^3 \pcc$. While there is no significant trend in the HCN/$^{12}$CO ratio, N$_2$H$^+$/HCN is strikingly linear with the DGF. This is due to the behaviour shown in Figure \ref{fig:peak}, where lines increase sharply in intensity when the peak line-of-sight density exceeds a threshold value, and more shallowly beyond this. The line ratio therefore effectively traces the fraction of pixels containing high-density gas, and thus the DGF, independently of the environmental conditions; the $\gamma = 1$ and $30$ simulations both fall on the same linear relationship.

We note that this strong connection between line emission and the mass reservoir in part explains the {\it lack} of correlation between line emission and the SFR. As noted above, the accreted sink mass tracks the mass of dense gas quite closely, with an offset of around a Myr, but the SFR by our definition is the {\it derivative} of the sink mass, which fluctuates quite significantly on the $100 \kyr$ timescales we measure it over, as shown in Figure \ref{fig:sfr}. Observational definitions of the SFR measure it as an integrated quantity over at least a few Myr, smoothing out these fluctuations, and likely contributing towards the much tighter linear correlations found between dense gas and the SFR on both local \citep{lada2010} and extragalactic \citep{gao2004} scales. Our simulations only cover around a Myr of star-forming activity\footnote{Following their evolution further would require the inclusion of stellar feedback, as the total accreted sink mass reaches the point where significant disruption by photoionisation is expected \citep{whitworth1979,walch2012}.}, so we cannot meaningfully assess changes in the Myr-averaged SFR; simulations of larger, kpc-sized regions of galactic disc \citep[e.g.][]{panessa2023} would be better suited to compare line emission properties with an SFR more closely related to the observational definition.

{We have focused so far on the $J=1-0$ transitions of the molecules in question, as these are the most frequently observed and (for HCN) form the basis of the Gao-Solomon relationship. There have been suggestions \citep[e.g][]{rybak2026} that higher-$J$ HCN transitions may follow a similar luminosity-SFR relationship; the higher upper-level energies of these lines would tend to make them more specific tracers of dense gas than their lower-$J$ counterparts. Figure \ref{fig:alpha32} shows the intensity-column density relationships for the HCN and N$_2$H$^+$ $J=3-2$ transitions from our simulations. For both molecules and both values of $\gamma$, the higher-$J$ transition produces negligible emission below a column density of $10^{22} \pcs$. This suggests that the HCN $J=3-2$ line, unlike the $1-0$, does act as a direct tracer of the reservoir of star-forming gas. However, extragalactic detections of this line are typically restricted to extreme systems such as central starbursts \citep[e.g.][]{behrens2026}, where disentangling the various possible excitation mechanisms may make a simple luminosity-dense gas-SFR relationship difficult to establish.}

\subsection{The radiation field in star-forming clouds}
\label{sec:radfield}

The comparison of our simulated $^{12}$CO line intensities as a function of column density to the Perseus data from \citet{tafalla2021} in Figure \ref{fig:alpha} suggests that the increased UV background of our $\gamma = 30$ model may better reproduce the observations at high column densities, by raising the temperature of the moderate-density gas where this emission originates. While the assumption of a Solar neighbourhood UV background is almost ubiquitous in studies of star formation, the Sun is much further removed from sites of active high-mass star formation than most molecular clouds \citep{zucker2023}. The average star-forming cloud may experience a significantly stronger external UV field than is generally assumed to be the case.

An additional piece of evidence supporting this is the relative behaviour of the HCN and HCO$^+$ lines. We have noted previously that under `standard' local ISM conditions, simulations predict that HCO$^+$ traces higher-density gas than HCN \citep{priestley2023a,priestley2024}, whereas there is no sign of this occurring in observations of nearby molecular clouds \citep{pety2017,tafalla2021,tafalla2023}. More generally, the filling factors of HCN and HCO$^+$ emission are found to be close-to-equal in environments ranging from giant molecular filaments (GMFs; \citealt{wang2020,feher2025}), high-mass star-forming regions \citep{barnes2020}, molecular clouds in the Andromeda \citep{forbrich2023} and Whirlpool \citep{stuber2025b} galaxies, and outer regions of the Milky Way \citep{patra2022}.

We compare line emission filling factors from our simulations to observed values from \citet{feher2025} and \citet{barnes2020} in Figure \ref{fig:fillfac}; these are calculated using a $1 \kel \kms$ detection limit comparable to the sensitivity of the observations, and normalised to the area covered by $^{12}$CO emission ($J=1-0$ for the simulations and W49, $J=3-2$ for the GMFs). For $\gamma = 1$, the HCN filling factor is notably higher than that of HCO$^+$, while it is significantly reduced for $\gamma = 30$. The filling factors of the HCO$^+$ and N$_2$H$^+$ lines are almost unchanged. The overall pattern is in much better agreement with the observational data for $\gamma = 30$. While this might be unsurprising in W49, a high-mass star-forming region, it appears that more quiescent regions are also better-characterised by an ambient UV field well above the Solar neighbourhood value. As a higher UV field may affect many other cloud properties, such as the initial mass function of the stars it forms \citep{cusack2025}, we suggest that theoretical studies of star formation take this possibility into account.

\section{Conclusions}

We have performed radiative transfer modelling of line emission from simulated molecular clouds, incorporating a realistic treatment of ISM thermodynamics and time-dependent chemical evolution to obtain self-consistent molecular abundances. We find that the HCN $J=1-0$ line does not trace the reservoir of star-forming gas, showing no variation in luminosity over several orders of magnitude in SFR. The N$_2$H$^+$ $J=1-0$ transition does increase with the SFR, but this manifests as a significant jump in luminosity following the onset of star formation, rather than a directly-linear relationship. The N$_2$H$^+$/HCN line ratio is very strongly correlated with the dense gas fraction of the clouds.

This is entirely consistent with resolved observational studies of nearby molecular clouds, but raises the question of why the HCN luminosity is found to correlate linearly with the SFR on larger extragalactic scales \citep{gao2004,neumann2025}. We suggest that the HCN-SFR relationship results from these extragalactic observations consisting of a large ensemble of individual molecular clouds in each resolution element, which on average contain the same fraction of dense, star-forming gas independently of their environment. The HCN line does not trace this star formation reservoir, which instead is most reliably associated with N$_2$H$^+$ due to unique features of its molecular chemistry.

\section*{Acknowledgements}

FDP, PCC, SER and OF acknowledge the support of a consolidated grant (ST/W000830/1) from the UK Science and Technology Facilities Council (STFC). FDP and PCC additionally acknowledge support from an STFC Small Award (UKRI1187). SCOG and RSK acknowledge funding from the European Research Council (ERC) via the ERC Synergy Grant “ECOGAL-Understanding our Galactic ecosystem: From the disk of the Milky Way to the formation sites of stars and planets” (project ID 855130), from the Heidelberg Cluster of Excellence (EXC 2181 - 390900948) “STRUCTURES: A unifying approach to emergent phenomena in the physical world, mathematics, and complex data”, funded by the German Excellence Strategy, and from the German Ministry for Economic Affairs and Climate Action in project ``MAINN'' (funding ID 50OO2206). The team in Heidelberg also thanks for computing resources provided by {\em The L\"{a}nd} through bwHPC and DFG through grant INST 35/1134-1 FUGG and for data storage at SDS@hd through grant INST 35/1314-1 FUGG. This research was undertaken using the supercomputing facilities at Cardiff University operated by Advanced Research Computing at Cardiff (ARCCA) on behalf of the Cardiff Supercomputing Facility and the Supercomputing Wales (SCW) project. We acknowledge the support of the latter, which is part-funded by the European Regional Development Fund (ERDF) via the Welsh Government.

\section*{Data Availability}
The data underlying this article will be shared on request.

\bibliographystyle{mnras}
\bibliography{sfrate}

\appendix

\section{Modifications to the NEATH chemical model}
\label{sec:neath}

\begin{figure*}
  \centering
  \includegraphics[width=0.67\columnwidth]{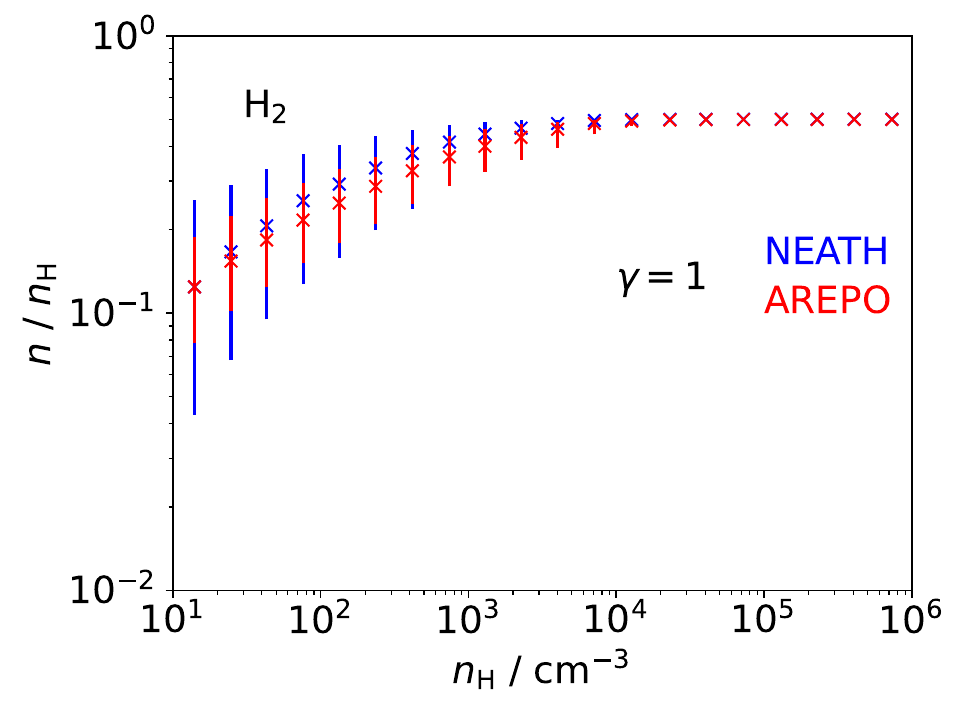}
  \includegraphics[width=0.67\columnwidth]{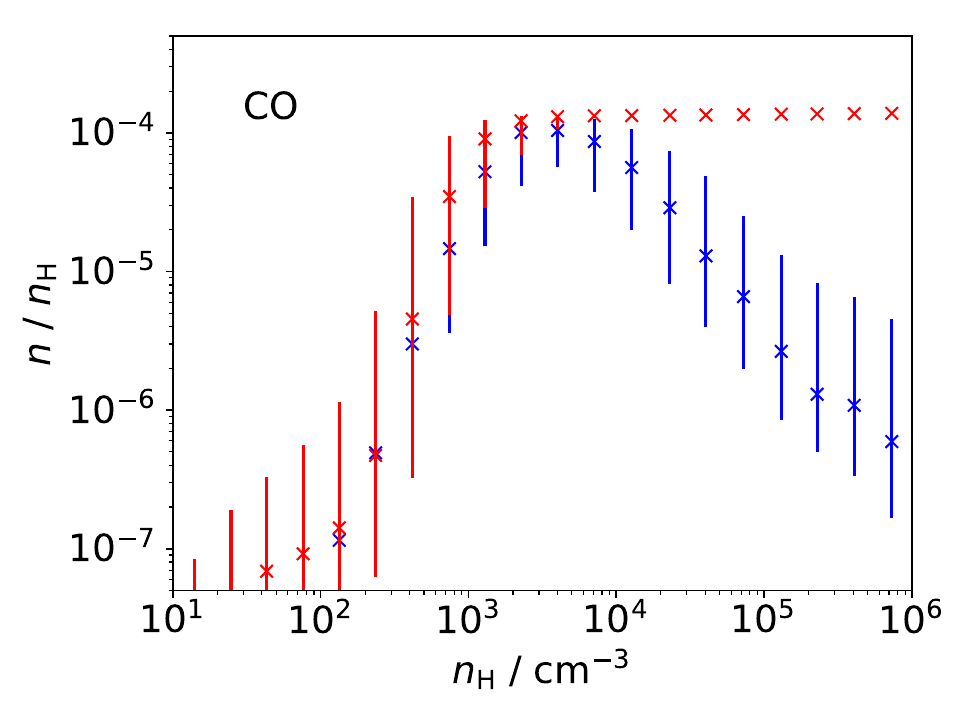}
  \includegraphics[width=0.67\columnwidth]{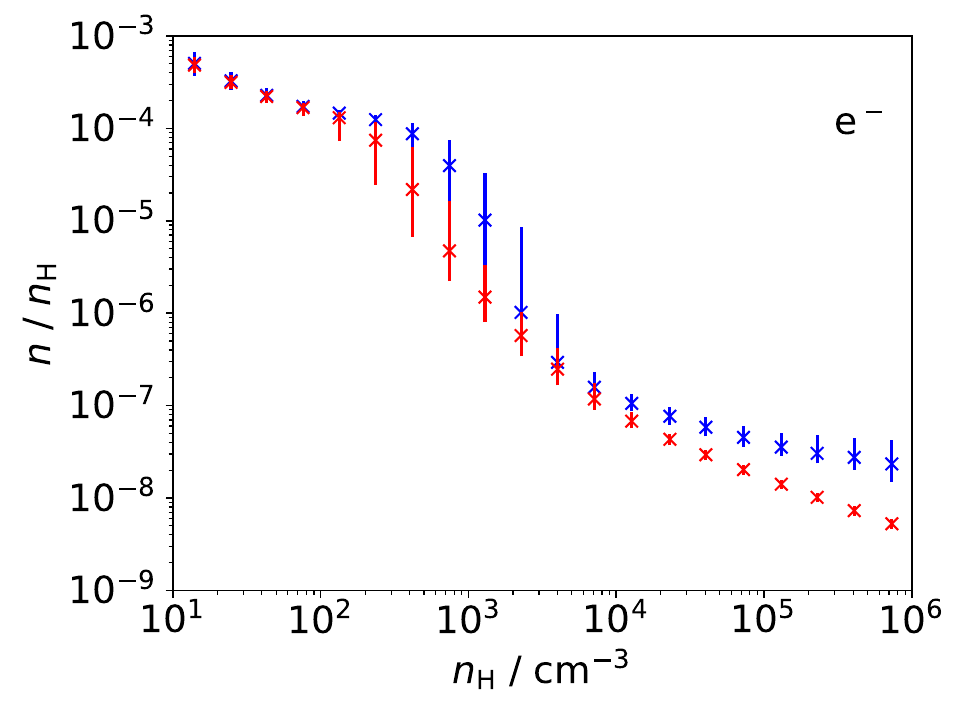}\\
  \includegraphics[width=0.67\columnwidth]{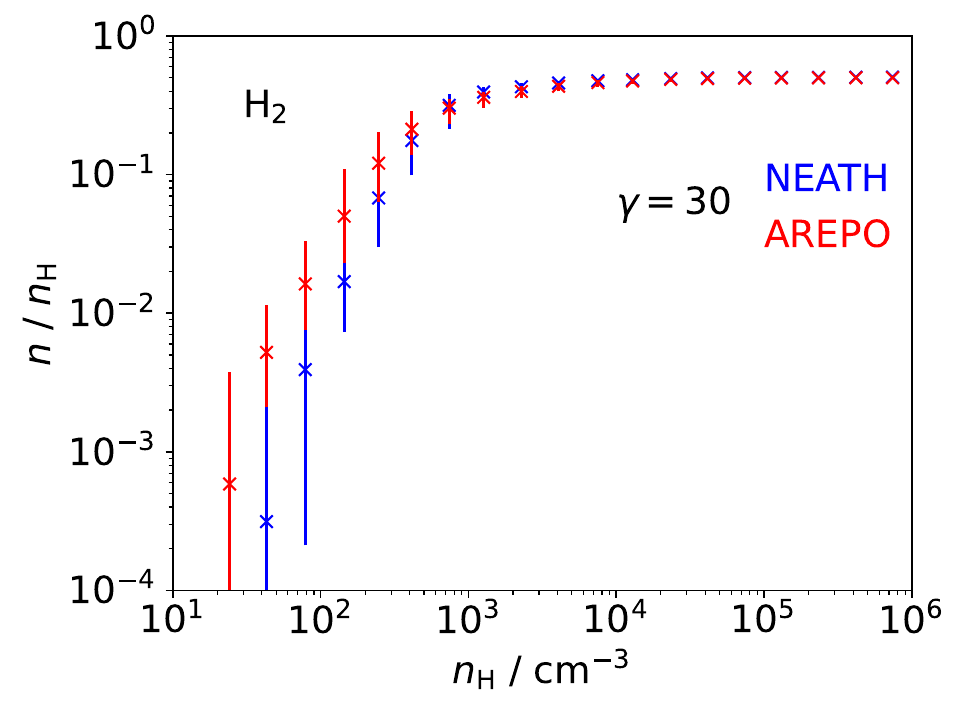}
  \includegraphics[width=0.67\columnwidth]{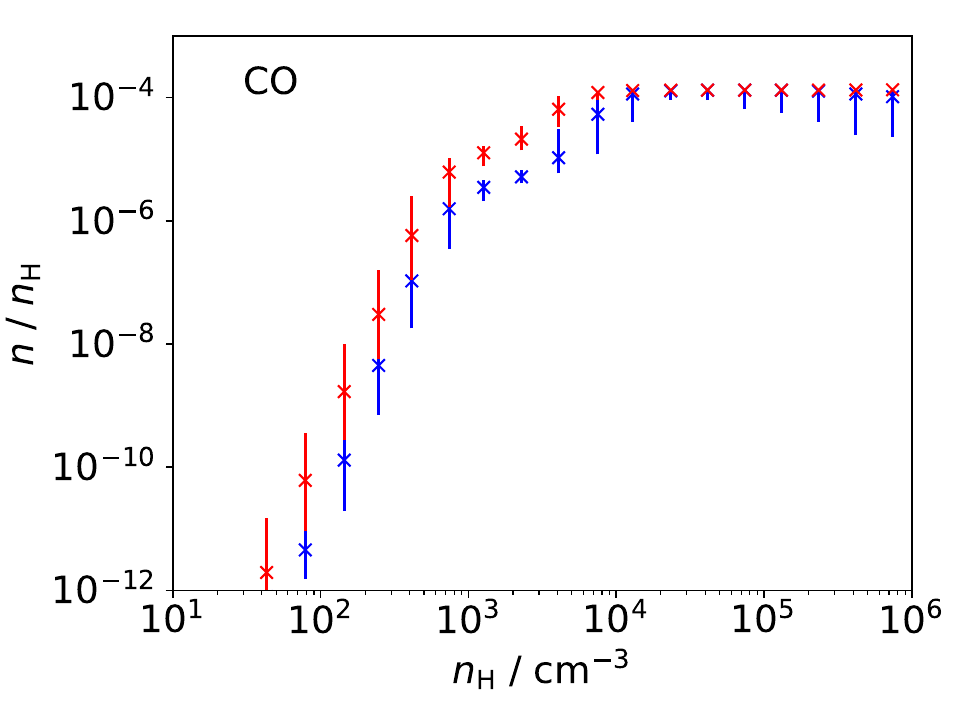}
  \includegraphics[width=0.67\columnwidth]{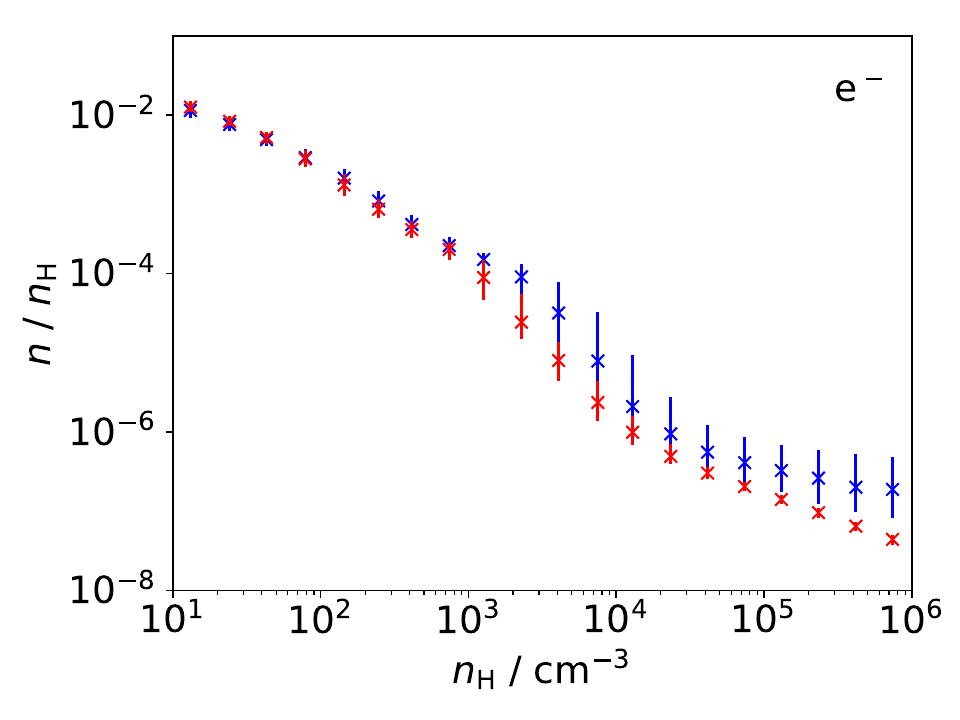}
  \caption{Average abundances as a function of $\nh$ for the NEATH (blue) and {\sc arepo} (red) chemical models: H$_2$ (left), CO (centre), and electrons (right). The top and bottom rows show results for the $\gamma = 1$ and $30$ simulations respectively. The median values are shown as crosses, with the bars indicating the 16th/84th percentiles.}
  \label{fig:neath}
\end{figure*}

The original NEATH chemical model described in \citet{priestley2023b} was calibrated to reproduce the H$_2$ and CO abundances of the internal {\sc arepo} network under standard Solar neighbourhood conditions ($G = 1.7 \, {\rm G_0}$, $\zeta = 10^{-16} \, {\rm s^{-1}}$). At higher CRIRs, the agreement between the internal and post-processed abundances was found to break down. The primary cause is the lack of grain-assisted recombination in the original model, resulting in much higher abundances of highly-reactive positive ions (particularly He$^+$) and electrons \citep{gong2017}. We have accordingly implemented grain-assisted recombination reactions for H, He, C, Mg, S and Si, following \citet{weingartner2001} with modifications as described in \citet{gong2017} and \citet{hunter2023}. This implementation does not conserve charge: positive ions react with dust grains to become neutral atoms, without any corresponding loss of free electrons. We have therefore modified {\sc uclchem} so that the electron abundance is reset to be equal to the sum of all positively-charged ions after every network integration. We also set the rate equation for the electron abundance to the sum of ion production/loss channels, to mitigate against rapid changes occuring within a single solver call. The updated version of the code is freely available on GitHub (\url{https://github.com/fpriestley/neath/}).

The standard version of {\sc uclchem} does include interactions between positive ions (other than H and He) and dust grains, which are assumed to result in neutralised and hydrogenated ice species (e.g. the freeze-out of C$^+$ produces CH$_4$ ice). As this process also does not conserve charge, an electron `freeze-out' process is included in the standard reaction network, adopting the \citet{rawlings1992} rate for a positively-charged ion with the mass of hydrogen. This approximately balances the loss of positive species for low CRIRs, but fails at higher values; as our updated version of {\sc uclchem} now explicitly conserves charge, we have deactivated the electron freeze-out reaction. We retain the freeze-out of positive ions, but the rates of these reactions are orders of magnitude lower than grain-assisted recombinations, so they have a negligible effect on the chemistry. We additionally modify the \citet{rawlings1992} rates for all freeze-out reactions with an $\exp\left(- T / 10^3 \kel \right)$ suppression factor, taken from the \citet{dejong1977} prescription for grain-surface H$_2$ formation, to prevent ice mantles forming at unrealistic temperatures.\footnote{Note that $T$ is the gas temperature; the dust temperature is not explicitly a parameter in the version of {\sc uclchem} on which NEATH is based. Grain-surface reactions assume the dust and gas temperatures are equal, but the network we use here does not include any of these processes.}

In Figure \ref{fig:neath}, we compare the NEATH and {\sc arepo} abundances of H$_2$, CO and electrons for the $\gamma = 1$ and $30$ simulations {at the simulation endpoints}. The decline in the NEATH CO abundance at high density is due to freeze-out, which is not modelled in the {\sc arepo} network. Similarly, the higher NEATH electron abundances at moderate densities are due to the neglect of neutral carbon self-shielding (only self-shielding of H$_2$ and CO are currently modelled), while at high densities the excess is due to the additional heavy elements in the network (S, Si and Mg; {\sc arepo} only has Si). Finally, between $10^3-10^4 \pcc$ in the $\gamma = 30$ simulation, the NEATH CO abundances are a factor of $3-5$ lower than the {\sc arepo} values. This can be traced to an underabundance of OH compared to the OH$_x$ proxy species in the \citet{gong2017} network. The latter is assumed to form instantaneously by the pseudoreaction between H$_3^+$ and O, whereas in the full network, there are multiple intermediate states (OH$^+$ and H$_2$O$^+$) which are vulnerable to destruction by reactions with other species. Overall, the two networks are in good agreement given the uncertainties inherent in astrochemical modelling, and the main purpose of this benchmarking has been achieved: were we to run MHD simulations with the full chemical network rather than the reduced one, we would not expect to see appreciably different physical results.

\section{Simulation properties}
\label{sec:data}

\begin{table*}
  \centering
  \caption{Physical properties and line intensities for the five snapshots of the $\gamma = 1$ and $30$ simulations used in the analysis.}
  \begin{tabular}{cccccccccc}
    \hline
    & Myr & \multicolumn{3}{c}{M$_\odot$} & M$_\odot$ Myr$^{-1}$ & \multicolumn{4}{c}{K km s$^{-1}$} \\
    $\gamma$ & $t$ & $M_{\rm mid}$ & $M_{\rm dense}$ & $M_{\rm acc}$ & SFR & $I_{\rm CO}$ & $I_{\rm HCN}$ & $I_{\rm N_2H^+}$ & $I_{\rm HCO^+}$ \\
    \hline
    & $3.48$ & $2066$ & $22$ & $0$ & $0.0$ & $0.879$ & $0.268$ & $0.0048$ & $0.091$ \\
    & $4.79$ & $3463$ & $190$ & $17$ & $14.7$ & $1.404$ & $0.297$ & $0.0286$ & $0.202$ \\
    $1$ & $5.01$ & $3385$ & $225$ & $23$ & $56.4$ & $1.403$ & $0.288$ & $0.0314$ & $0.214$ \\
    & $5.22$ & $3320$ & $222$ & $46$ & $130$ & $1.417$ & $0.281$ & $0.0344$ & $0.224$ \\
    & $5.44$ & $3049$ & $212$ & $85$ & $219$ & $1.404$ & $0.276$ & $0.0377$ & $0.232$ \\
    \hline
    & $6.09$ & $453$ & $14$ & $0$ & $0.0$ & $0.134$ & $0.0070$ & $0.00024$ & $0.0182$ \\
    & $6.75$ & $539$ & $83$ & $14$ & $75.1$ & $0.146$ & $0.0100$ & $0.00377$ & $0.0231$ \\
    $30$ & $6.92$ & $535$ & $126$ & $27$ & $175$ & $0.142$ & $0.0108$ & $0.00613$ & $0.0266$ \\
    & $7.09$ & $495$ & $133$ & $80$ & $298$ & $0.137$ & $0.0112$ & $0.00774$ & $0.0289$ \\
    & $7.27$ & $453$ & $138$ & $135$ & $461$ & $0.132$ & $0.0110$ & $0.00750$ & $0.0276$ \\
    \hline
  \end{tabular}
  \label{tab:data}
\end{table*}

The data used to investigate correlations between our simulations' physical and line emission properties are given in Table \ref{tab:data}. We performed radiative transfer modelling for five snapshots from each of the $\gamma = 1$ and $30$ simulations, chosen to cover a representative range of SFRs. At each snapshot, we calculate the mass in the simulation above density thresholds of $10^3 \pcc$ ($M_{\rm mid}$) and $10^4 \pcc$ ($M_{\rm dense}$), and the total mass of sink particles ($M_{\rm acc}$). The SFR at each snapshot is calculated as the change in $M_{\rm acc}$ between the preceding and following snapshots, divided by the $88 \kyr$ interval between them. Line intensities are average values within a radius of $16.2 \pc$ from the centre for the simulations seen edge-on; this aperture size was chosen to encompass all significant line emission for all snapshots and both $\gamma$ values.

\bsp	
\label{lastpage}
\end{document}